\documentclass[twocolumn,preprintnumbers,amsmath,amssymb]{revtex4}
\usepackage{graphicx}
\usepackage{amsmath}
\usepackage{amssymb}
\usepackage{amsthm}
\usepackage{amsfonts}
\usepackage{enumerate}
\usepackage{overpic}
\usepackage{centernot}
\usepackage{tikz}

\begin{document}

\title{Universal topological quantum computation with strongly correlated Majorana edge modes}

 \author{Ye-Min Zhan$^{1,2,3}$}
  \thanks{These two authors contribute equally.}
 \author{Yu-Ge Chen$^4$}
 \thanks{These two authors contribute equally.}
 \author{Bin Chen$^5$}
 \author{Ziqiang Wang$^6$}
 \author{ Yue Yu$^{1,2,3}$}
 \thanks{Correspondence to: yuyue@fudan.edu.cn}
 \author{Xi Luo$^5$}
\thanks{Correspondence to: xiluo@usst.edu.cn}

 \affiliation {1. State Key Laboratory of Surface Physics, Fudan University, Shanghai 200433,
 	China\\
 	2.Center for Field
 	Theory and Particle Physics, Department of Physics, Fudan University, Shanghai 200433,
 	China\\
 	3. Collaborative Innovation Center of Advanced Microstructures, Nanjing 210093, China\\
	4. Institute of Physics, Chinese Academy of Sciences, Beijing 100190, China\\
	5. College of Science, University of Shanghai for Science and Technology, Shanghai 200093, China\\
        6. Department of Physics, Boston College, Chestnut Hill, MA 02467, USA }
\date{\today}

\begin{abstract}  Majorana-based quantum gates are not complete for performing universal topological quantum computation while Fibonacci-based gates are difficult to be realized electronically and hardly coincide with the conventional quantum circuit models.  In Ref. \cite{hukane}, it has been shown that a strongly correlated Majorana edge mode in a chiral topological superconductor can be decomposed into a Fibobacci anyon $\tau$ and a thermal operator anyon $\varepsilon$ in  the tricritical Ising model. The deconfinement of $\tau$ and $\varepsilon$ via the interaction between the fermion modes yields the anyon  {collisions} and gives the braiding of either $\tau$ or $\varepsilon$. With these braidings, the complete members  {of} a set of universal gates,  the Pauli gates, the Hadamard gate and extra phase gates for 1-qubit as well as  controlled-not gate for 2-qubits, are topologically assembled. Encoding quantum information and reading out the computation results can be carried out through electric signals. With the sparse-dense mixed encodings, we set up the quantum circuit  {where the controlled-not gate turns out  { to be} a probabilistic gate} and design the corresponding devices with thin films of the chiral topological superconductor. As an example of the universal topological quantum computing,  we show the application to Shor's integer factorization algorithm.
 \end{abstract}

\maketitle

\section{ Introduction}

The idea of quantum computation can be traced back to the 1980s by  {Manin, Benioff, and Feynman et al.}, who tried to simulate the quantum world by coherent quantum states and quantum models \cite{Manin1980,Benioff1980,Feynman1982}. In the  {19}90s, Shor's integer factorization algorithm provides a theoretical example of the applications of quantum computation, which has an enormous advantage over its classical counterpart \cite{Shor1995}. Nowadays, the realizations and applications of quantum computation have become one of the most frontier topics in the developments of quantum technology \cite{Gibney2019}. In the attempts of constructing the prototypes of quantum computers, there have been proposals based on the entangled states of photons \cite{Zhong2020}, superconducting circuits \cite{Clarke2008}, bound states in cold atoms \cite{Cirac1995} or optical lattices \cite{Mlmer2020}, quantum wells \cite{Ivscdy2019},  quantum dot spins \cite{Imamoglu1999}, single spins in diamonds \cite{Neumann2008}, Bose-Einstein condensates \cite{Anderlini2007}, rare-earth atoms \cite{Ohlsson2002}, and carbon nanospheres \cite{Nscfrscdi2016}. Though quantum computing provides promising applications in quantum simulation, data mining, machine learning, deciphering, artificial intelligence, etc., the road to realizing  {the} quantum computation remains tough. In the theoretical side,  {besides predicting material realizations, another serious challenge lies} in making the quantum computation universal, i.e., to realize a protocol for universal quantum computation based on quantum systems and quantum logic gates such as the controlled-not (CNOT) gate. While in the experimental side,  {one serious challenge is} the decoherence between the quantum states and the environment which will cause errors. For example, to create an error-tolerant logic gate for a singlet logic qubit, more than 150 physical logic gates are needed when processing in Steane codes \cite{Steane2004,Shu}.

\subsection{Topological Quantum Computation}

Recent progresses on topological quantum states of matter have shed new light on overcoming these obstacles. The concept of topological quantum computation (TQC) was first proposed by Kitaev \cite{K1,TQCR}, which is believed to be fault-tolerant due to the protection of the topological gap between the system and the environment. The computational data is stored non-locally by the topological excitations, namely, anyons. In the TQC, the initial data is prepared by coherent many-particle anyon states. The braidings between anyons generate unitary transformations that correspond to logic gates and the readout of the results are achieved by fusing the anyons into observable quantities \cite{Rowell}.

The concept of anyon was proposed by Leinaas and Myrheim \cite{Leinaas1977}, and Wilczek \cite{Wilczek1982}.  Laughlin pointed out that the quasi-particles with fractional charge  in the fractional quantum Hall effects (FQHE) could be anyons. In the $\nu=5/2$ even denominator FQHE, Moore and Read  {noticed} that the non-Abelian anyons are of Ising type \cite{MR,will}. Later, Freedman et al. proved that to achieve universal quantum computation, the topological quantum gates with these Ising type non-Abelian anyons alone are not enough, and additional non-topological phase gates, e.g., the $\frac\pi{8}$ gate, are needed \cite{Freedman2003}. Among the possible non-Abelian anyon systems, the Fibonacci anyon is the simplest case for achieving universal TQC \cite{TQCR}, which is likely to exist in FQHE with $\nu=12/5$ \cite{rr,Nayak2008},  but there are  substantial uncertainties.  A recent study reported the possibility that the Fibonacci anyon appears in the $\nu=\frac{2}3$ FQHE, appropriately proximitized by a superconductor \cite{mong,vaezi}. But this requires the survival of the superconductivity in a strong magnetic field.

Besides FQHE,  {Kitaev proposed Majorana bound states at the two ends of topological superconducting quantum wires \cite{K3}.} Two Majorana zero modes (MZMs) at the two ends have a $\pi/2$ phase shift. They correspond to the "real" and "imaginary" part of a non-local fermion. Except for the phase difference, the higher dimensional non-Abelian representations of the braid group of the MZMs are equivalent to the Ising type  {anyons'} \cite{NW,Inv,prox1}.   Some quantum gates for TQC can be achieved by  {exchanging the Majorana bound states in so-called T-junctions} \cite{prox1}.  Unfortunately, the MZMs also suffer from lacking of topological phase gates for universal TQC.  Even so,  Kitaev's seminal work has stimulated the physicists' enthusiasm for finding MZMs. Using the proximity effect of the $s$-wave superconductor, Fu and Kane proposed a theory for MZM in the superconducting vortex at the interface between an $s$-wave superconductor and a topological insulator \cite{fukane}. There are other proposals for MZMs in semiconductor heterostructures \cite{Sau2010}, semiconductor-superconductor heterostructures \cite{Das Sarma2010}, superconductor/2D-topological-insulator/ferromagnetic-insulator hybrid system \cite{Luo1}, and so on. Besides the progresses on the theories, there are also lots of experimental reports on MZMs. Midgap states at zero bias voltage were observed in indium antimonide nanowires in a magnetic field \cite{Mourik2012,Deng2012,ADas,Chur,Deng}. Although the results are consistent with the existence of MZMs \cite{Das Sarma2010}, the zero energy states can also be explained by other non-topological trivial bound states \cite{Lee2013}.  Possible MZMs were observed at the ferromagnetic atomic chains on a superconductor \cite{Nadj-Perge2014}, but there can be other explanations besides MZM. A strong indirect evidence for MZM was observed in Bi$_2$Te$_3$/NbSe$_2$ heterostructure \cite{Jia1,Sun2016}. A $2e^2/h$ conductance was observed in InSb nanowire, which is believed to be an evidence for MZM \cite{H.Zhang2018},  but this result was retracted. And the authors' latest data show more possible interpretations besides MZM \cite{H.Zhang}. Recently, the evidences of MZMs are reported in iron-based superconductors, such as FeTe$_{0.55}$Se$_{0.45}$ \cite{Kong2019,Chen2020,wang2020}, (Li$_{0.84}$Fe$_{0.16}$)OHFeSe \cite{Chen2019}, LiFeAs \cite{P. Zhang2019}, and CaKFe$_4$As$_4$ \cite{W. Liu2020}, as well as the Majorana
vortex states in iron-based superconducting nanowires \cite{LiuX},  {and MZM are also reported in atomically Fe-based Yu-Shiba-Rusinov chains \cite{ysr1,ysr2,ysr3},} which provide new materials for creating Ising type topological quantum computer \cite{FCZ}.

The MZMs in FQHE with $\nu=5/2$ and Kitaev's model are closely related to the vortex bound states and chiral Majorana edge modes ($\chi$MEMs) in the $p$-wave superconductor \cite{RG2000}.  The interior of a vortex can be regarded as vacuum with a $\chi$MEM moving along the edge of the vortex, and when the radius of the vortex goes to zero, the $\chi$MEM reduces to the MZM, i.e., this provides a duality between MZM and MEM. Therefore, the $p$-wave topological superconductor is an important candidate for realizing the TQC. Although the material realization of the $p$-wave superconductor has not been found yet, for example, Sr$_2$RuO$_4$, which remains controversial \cite{Ishida1998,Pustogow2019}, the materials with effective $p$-wave superconductivity are possible to construct.  One possible scenario is by using the $s$-wave superconductor/quantum anomalous Hall heterostructure, and the proximity effect will induce a chiral topological superconducting ($\chi$TSC) phase \cite{X.-L. Qi2010,BL}. Furthermore, the braiding of the MEMs are carried automatically when propagating along the edge, and the $e^2/2h$ conductance is believed to be a smoking gun signal for the existence of MEM \cite{BL}. Unfortunately, the experimental evidences of this scenario remain controversial \cite{Science,X.-G. Wen2018,science1}. The reason is that in this heterostructure, there can be residual metallic states in the quantum anomalous Hall substrate, which can also explain the half quantized conductance \cite{X.-G. Wen2018}. And the MZMs in the vortices of the superconducting area will dramatically change the results of the braiding \cite{Halperin2006, Kitaev2006}.

We have briefly review the recent progresses in the TQC, especially for those based on Majorana objects. For more recent subject review, see \cite{ayu}. We can classify these Majorana-based TQC proposals into two approaches to the non-Abelian statistics: (1) For one species Majorana fermion system,  the many-particle wave functions of Majorana fermions obey the Abelian statistics. Other degrees of freedom such as the vortices must exist to have non-Abelian statistics, e.g, in the $\nu=5/2$ fractional quantum Hall states and  the Ising model \cite{MR,NW} as well as the edge chiral vortices approach \cite{Bee}. This kind of Majorana fermion systems can do the TQC but not universal by braiding solely.  (2) For two species Majorana fermion model which is we consider here, no further degrees of freedom such as vortices are necessary. The two species Majorana fermions may be non-Abelian when the fermion number conservation is reduced to the fermion parity (FP) conservation, e.g., in superconductors. This approach has been used in \cite{NW,Inv,K3,prox1, X.-L. Qi2010,BL}. This kind of Majorana fermion systems  {also} cannot be used to do  {universal} quantum computation by braiding solely. In Appendix \ref{A}, we give more explanations to these two approaches for readers' convenience.

\subsection{The Goal and Main Results of This Work}

We already know that there are TQC processes with two types of non-Abelian anyons. The Fibonacci anyon-based one is universal but restricted by realistic materials. The algorithm is also different from the conventional quantum circuit models. On the other hand, the Majorana fermion-based one is relevant to practical physical systems and the quantum circuit models, but is not universal by braiding solely.  {Furthermore, there should be no other low-energy fermionic or apparently even no bosonic states in the system to avoid the Majorana modes overlapping effectively, which is another obstacle for  Majorana fermion-based TQC \cite{prb85,njp24}.} The main goal of this article is to design quantum gates for the universal TQC consistent with the quantum models,  which combines the advantages of the $\chi$MEMs and the Fibonacci-type anyons.

In a recent work concerns Fibonacci topological superconductor \cite{hukane}, the authors showed that, in 7-layers of $\chi$TSC, the interacting $\chi$MEMs $\gamma_a~ (a=1,...,7)$ whose conformal dimension is $\frac{1}2$ may be thought as composite objects: $\gamma_a=\tau_a\varepsilon_a$ where $\tau$ is the Fibonacci anyon with the conformal dimension $\frac{2}5$ and $\varepsilon$ is a thermal operator with the conformal dimension $\frac1{10}$ in the tricritical Ising (TCI) model. The validity of such a composition is guaranteed by the coset factorization of the conformal field theory (CFT): SO(7)$_1=(G_2)_1\times$ TCI \cite{Sha}.  (For readers' convenience, we briefly list the basic facts of $G_2$ in Appendix \ref{B}.)

 The components of the MEMs can be delocalized when two MEMs with opposite chirality interact through a special interacting potential \cite{hukane}. We here will analyze this interaction and show that it can be expressed as the interacting potential between charged fermions which consist of the MEMs when they meet in the interacting domains.  Furthermore,  the interaction between two propagating MEMs yields the collisions of either two $\tau$ or two $\varepsilon$. When the collided $\tau$ and $\varepsilon$ are recombined, the braidings of anyons are achieved, similar to the Laughlin anyons collision \cite{anyoncoll}. \\
 
With these anyons and their braidings, we obtain:\\
 
 (1) These braidings {cause} the fractional statistical angles $\theta$ and then give the topological $\theta$-phase gates with $\theta=\frac{\pi}4, \frac{\pi}{10}$ and $\frac{2\pi}5$. With the $\frac{\pi}4$-phase gate and the braiding gate by braiding two MEMs in different pairs of the $\chi$MEMs \cite{BL},  the Pauli-$X,Y, Z$ gates, the $H$ gate,  and the CNOT gate can be created. The other two phase gates, the $\frac\pi{10}$- and $\frac{2\pi}5$-phase gates, which are topological, can replace the $\frac{\pi}8$-phase gate. With the similar method in \cite{vatan}, we prove that all these gates form a set of universal gates. We then can design a universal TQC with quantum circuit model \cite{UTQC}.
 
 (2) We find that encoding quantum information and reading out the computation results can be carried out through electric signals. 
 
 (3) The quantum gates are generally dependent on the FP, the even or odd of the fermion number of the quantum state in the process, which is conserved and then determined by the initial qubits. For example, for a device set up by fixed physical elements of a phase gate, the 1-qubit initial state with odd FP gives a phase gate diag(1, i) while it is diag(-i, 1) for even FP.   A device gives the CNOT gate if the 2-qubits are of an odd FP but it does not give a CNOT for the FP even qubits and vice verse, which is discussed in more details.  
 
 (4) Therefore, for a given physical device, side measurements are needed in order to have the input state with correct FP for the designed gate.  The side measurement requires additional MEMs besides those in the computing state space. It is known that to encode the Majorana-based $n$ logic qubits, $2n+2$ Majorana bound states or MEMs are needed \cite{NW}. However,  with the side measurements, $4n$ MEMs are required to encode the $n$-qubits \cite{DSFN}. The former is usually called the dense encoding process while the latter is  {the} sparse encoding one \cite{DSFN}. Using the dense encoding only is not convenient to build a quantum circuit which is applied to well-known quantum algorithms, e.g., the quantum Fourier transformation and Shor's integer factorization algorithm. Thus, the sparse encoding will be used to build our quantum circuit. However, in the sparse encoding, one can only prepare qubits in superposition states, but no entanglement states with braiding only \cite{Bravyi}. For instance, with the sparse encoding, we cannot realize the CNOT gate from two 1-qubits. It has to be realized in the dense encoding. Thus,  our process is a mixed one with the sparse-dense encodings.   Notice that because  {of} the no entanglement rule, the FP measurements are needed. Therefore, our CNOT gate is still probabilistic, although the efficiency is improved.
 
 Using this process, we may realize a quantum circuit for Shor's algorithm.  Since there is one-to-one correspondence between the quantum gates and the designed devices with the $\chi$TSC thin films, we can design a universal TQC for the large integer primary factorization.
 
\subsection{Paper's Organization}
 
This paper is organized as follows:    In Sec. II,  we review a 7-layer TSC system and  the corresponding conformal field theory properties. In Sec. III,  we study the interaction between the right- and left-$\chi$MEMs ($R,L$-$\chi$MEMs) when they meet. In Sec. IV, we show the anyon exchange and braiding due to the interaction between the $\chi$MEMs. In Sec. V, we construct a set of universal topological gates which are used to encode a universal TQC. The corresponding physical devices are designed. In Sec. VI, we show that  the  computing results can be output by electric signals.  In Sec. VII,  we use these quantum gates to build the quantum circuit with the sparse-dense mixed encoding. We apply these designations to the quantum Fourier transformation, Toffoli gate which is the key element for an adder, and then the quantum circuit for Shor's algorithm. In Sec. VIII, we give the device designation of the quantum circuits by using the elements of the quantum gates which consist of the thin films of the $\chi$TSC and the interacting domains of the MEMs.  The last section devotes to our conclusions.  We have six appendices to support the results obtained in the main text.

\section{Multilayer $\chi$TSC systems}

 We consider multilayers of $\chi$TSC which are separated layer-by-layer by the trivial insulator. Spinless or spin polarized charged fermions are injected into the edges of the multilayers and each edge fermion is decomposed into two MEMs with different species: $\psi=\frac{1}{\sqrt{2}}(\gamma^{(1)}+i\gamma^{(2)})$ (See Appendix \ref{A}). Depending on the multilayer assignment, we label $\gamma_a$ the MEM into the $a$th layer. We do not explicitly index the species of the MEM because each of them freely runs in its own layer edge.  Thus, a $\chi$MEM on the edge of each layer is spatially separated from the other $\chi$MEMs. We denote the edge coordinate as $x$. The free $\chi$MEMs $\gamma_a^{R,L}(x)$ on the edges of an individual layer is described by the Hamiltonian
\begin{eqnarray}
H_{a}^{R,L}=\pm\frac{iv}2\gamma^{R,L}_a \frac\partial{\partial x} \gamma^{R,L}_a,
\end{eqnarray}
where $\pm$ are corresponding to the $right~(R)$- and $left~(L)$-chirality.    The $N$-layer $\chi$MEMs are described by the SO$(N)_1$ CFT \cite{hukane}. The fermion number $N$ of the system is not conserved but the even and odd of the fermion number, i.e., the FP $(-1)^N$, is conserved.

For our purpose, we take $N=7$.  It is known that if there are appropriate interactions between the MEMs (see below and Appendix \ref{C}),  the SO(7)$_1$ CFT can be factorized by the coset SO(7)$_1/$(G$_2)_1$. The central charge of the Wess-Zumino-Witten models of  a level $k$ affine Lie group $G$ is given by
$ c_G=\frac{k~{\rm dim} g}{k+g}$
 where dim$g$ is the dimension of the Lie algebra and $g$  {is} the dual Coxeter number. For SO(7)$_1$,  $c=\frac{21}{1+5}=\frac{7}2$ while $c=\frac{14}{1+4}=\frac{14}5$ for (G$_2)_1$. Therefore, the central charge of the coset SO(7)$_1/$(G$_2)_1$ is $\frac{7}2-\frac{14}5= \frac{7}{10}$. This means that the CFT of the Wess-Zumino-Witten model of  the coset SO(7)$_1/$(G$_2)_1$ is nothing but the TCI model \cite{Sha}.    The (G$_2)_1$ CFT has one type of  anyon: Fibonacci $\tau$ with  the conformal dimension $h_\tau=\frac{2}5$ while the SO(7)$_1/$(G$_2)_1$ CFT is equivalent to the TCI model where non-Abelian anyons, the thermal operators  $\varepsilon$ and $\varepsilon'$  have the conformal dimensions $h_\varepsilon=\frac{1}{10}$ and $h_{\varepsilon'}=\frac{3}5$, respectively \cite{Sha}. These non-Abelian anyons have the same quantum dimensions, i.e., $d_\tau=d_\varepsilon=d_{\varepsilon'}=\frac{1+\sqrt5}2$. Thus, as we have mentioned in  {\it Introduction}, the MEMs  can  {be} decomposed into $\gamma_a=\tau_a\varepsilon_a$ in terms of the coset construction due to the  adaptation of the conformal dimensions.

\section{ Interaction between R- and L-$\chi$MEMs for 7-layers}

  Formally, the seven free $\chi$MEMs Hamiltonian $H^{R,L}=\sum_{a=1}^7 H_a^{R,L}$ can be decomposed into $H^{R,L}_{G_2}+ H^{R,L}_{TCI}$. The explicit expressions of $H^{R,L}_{G_2}$ and $H^{R,L}_{TCI}$ can be found in  \cite{hukane} and are not important here, but we know that $[H^{R,L}_{TCI}, J^\alpha_{R,L}]=0$ where the current operators $J^\alpha_{R,L}~(\alpha=1,...,14)$ are defined by
$
 J^\alpha_{R,L}=\frac{1}2\sum_{a,b}\Xi^\alpha_{ab}\gamma^{R,L}_a\gamma^{R,L}_b
 $ for $\Xi^\alpha$ being the generators of the fundamental representation of $G_2$. Following Ref. \cite{hukane}, we consider the interaction between the $R$- and $L$-$\chi$MEMs as
 \begin{eqnarray}
 H_i=-\lambda\sum_{\alpha=1} ^{14}J^\alpha_RJ^\alpha_L. \label{int}
 \end{eqnarray}
 Using the quadratic Casimir operator of $G_2$ (see Eq. (\ref{qc})), $H_i$ is rewritten as \cite{hukane}
  \begin{eqnarray}
 H_i=-\frac{\lambda}3\sum_{a\ne b}\gamma^R_a\gamma^R_b\gamma^L_b\gamma^L_a-\frac{\lambda}3{\sum}'\gamma^R_a\gamma^R_b\gamma^L_c\gamma^L_d, \label{int1}
  \end{eqnarray}
 where $\sum'$ means the summation runs over the indices with $\epsilon_{abcdefg}\hat{f}_{efg}=-1$ (see Appendix \ref{B}).  If any two MEMs with different species  {but} the same chirality meet, they become a local charged fermion, say, $\psi^{R}_{ab}=\frac{1}{2}(\gamma^{R(1)}_a+i\gamma_b^{R(2)})$ and $\psi^{L}_{ab}=\frac{1}{2}(\gamma^{L(1)}_a+i\gamma_b^{L(2)})$ in Fig. \ref{fig1} (e). Since $i\gamma^{R(1)}_a\gamma^{R(2)}_b=2n^{R}_{ab}-1=2\tilde n^R_{ab}$ with the fermion number operator $n^R_{ab}=\psi^{R\dag}_{ab} \psi^R_{ab}$, for $\lambda>0$,  the interaction Hamiltonian becomes the interactions with a particle-hole symmetry between the charged fermions
  \begin{eqnarray}
 H_i&=&U\sum_{a\ne b}\tilde n^R_{ab}\tilde n^L_{ba}+U{\sum}' \tilde n^R_{ab}\tilde n^L_{cd}, \label{int2}
  \end{eqnarray}
where $U=\frac{4\lambda}3$.
If two MEMs with different species and chirality meet (Fig. \ref{fig1}(f)), the Hamiltonian can be expressed as  
\begin{eqnarray}
H_i=U\sum_{a\ne b}\tilde n^{RL}_{aa}\tilde n^{RL}_{bb}+U{\sum}' \tilde n^{RL}_{ad}\tilde n^{RL}_{bc},  
\end{eqnarray}
where $\tilde n^{RL}_{ad}=n^{RL}_{ad}-\frac{1}2$ for the fermion number operator $n^{RL}_{ad}=\psi^{RL\dag}_{ad} \psi^{RL}_{ad}$ with $\psi^{RL}_{ad}=\frac{1}{2}(\gamma^{R(1)}_a+i\gamma_d^{L(2)})$, and so on. Therefore, it is possible to realize the interactions, e.g., by introducing four narrow stripes of the $\chi$TSC sample from the edges of the thin films to a domain where the MEMs interact (See Appendix \ref{C}).  In reality, the coupling constant $\lambda$ may be dependent on the domains. But if the strengths of these coupling constants are of the same order, these differences are not relevant at the strong coupling fixed point. The $\tau$-anyon in the domain gains an energy gap  $\Delta\sim e^{-\pi v/2\lambda}$ for  $\lambda>0$ \cite{hukane}. This means that in the strong coupling region (see the green box in Figs. \ref{fig1}(a,b)), $\tau$ are reflected by this interaction potential while $\varepsilon$ are transmitted.  In this sense, the composite MEMs in the interacting domains are decomposed into spatially separated $\tau$ and $\varepsilon$.  Notice that the interactions $\gamma^R_a\gamma^R_b\gamma^L_b\gamma^L_d$ with $a\ne b\ne d$, etc will gap $\varepsilon$. Therefore, when introducing the MEMs to the interacting domains, one must avoid this type of interactions. The FP is also conserved after the interaction is switched on. \\

\section{Anyon braiding}

\begin{figure}
\centerline{\includegraphics[width=0.5\textwidth]{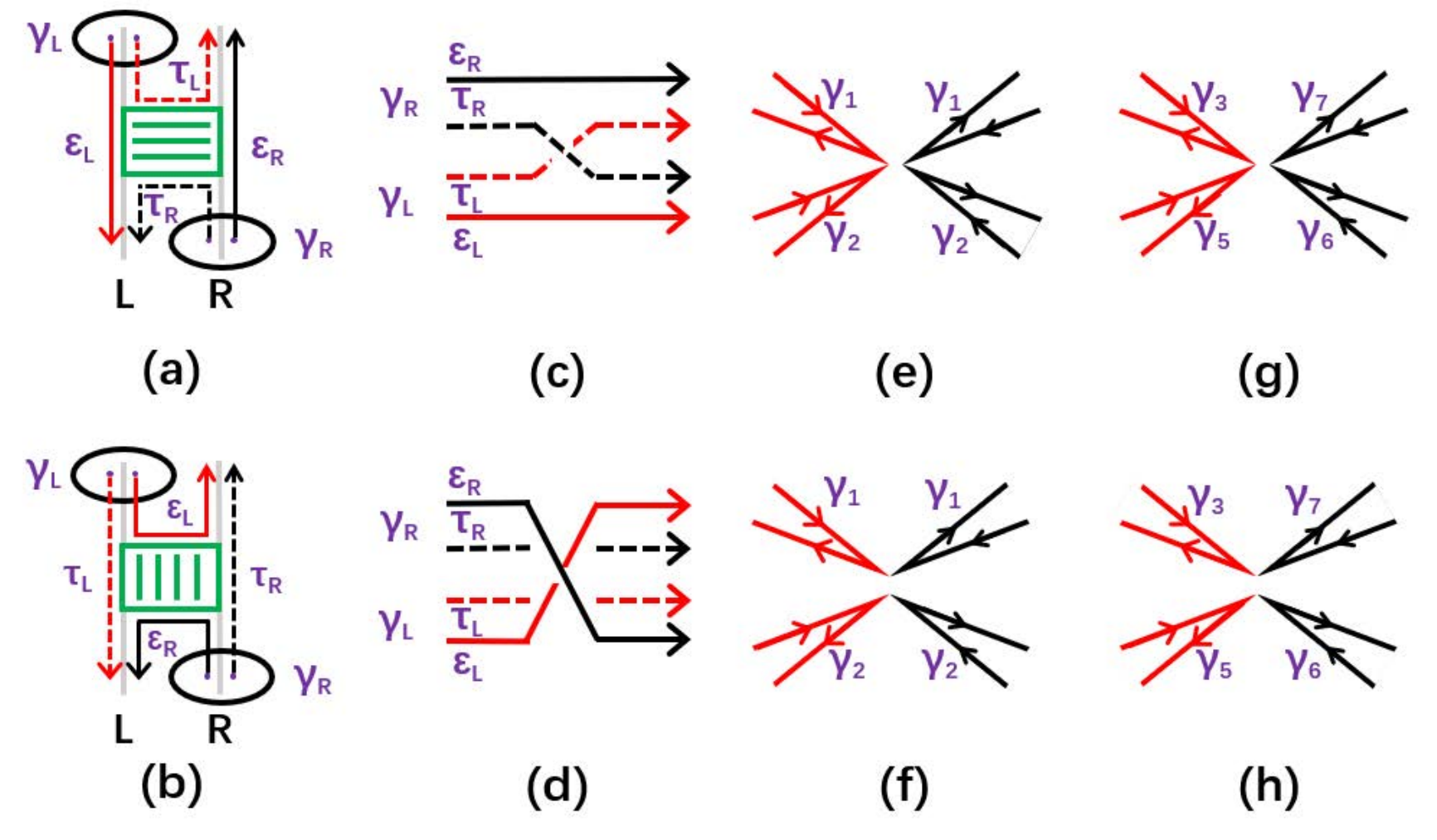}}
	\caption{(Color online)
		The illustrations of the anyon interactions, exchanges and braidings. The lines in the greens box in (a) and (b) stands for the interacting barrier for the $\tau$-anyons while the $\varepsilon$-anyons are free from the barrier. (a) The $\tau$s exchange. (b) The $\varepsilon$s exchange.  (c) and (d): The braiding world lines of the anyons that correspond to (a) and (b), respectively.  (e), (f), (g) and (h): Several examples of the interactions  with the solid line being the edge of the $\chi$TSC and the lower index in $\gamma$ being the layer index.  (e) $\tilde{n}^L_{12}\tilde{n}^R_{21}$, (f) $\tilde{n}^{LR}_{11}\tilde{n}^{LR}_{22}$, (g) $\tilde{n}^L_{35}\tilde{n}^R_{67}$, and (h) $\tilde{n}^{LR}_{37}\tilde{n}^{LR}_{56}$.		\label{fig1}	}
\end{figure}

 In Fig. \ref{fig1} (a) and (b), we depict two ways that a $R$-$\chi$MEM and a $L$-$\chi$MEM propagate to the interacting domain. In both cases, $\varepsilon_{R,L}$ are transmitted while $\tau_{R,L}$ are reflected by the interacting potential, but $\tau_R$ and $\tau_L$ exchange in  Fig. \ref{fig1} (a) while $\varepsilon_{R,L}$ exchange in Fig. \ref{fig1}(b). These anyon exchanges give rises the fractional statistics by braiding the anyons \cite{anyoncoll}. The world lines of the corresponding anyon braidings are showed in Fig. \ref{fig1} (c) and (d), respectively. The reason for why there are two types of collisions can be traced to the fact that we have two ways to write the four-Majorana fermion interaction in Eq. (\ref{int2}). Figs. \ref{fig1} (e) and (g) are the examples of the interacting processes corresponding to $\tau$'s braiding while Figs. \ref{fig1} (f) and (h) are to $\varepsilon$'s braiding.\\
 
 {For a CFT, the four-point correction function corresponding to the anyons collision can be calculated if we assume the $\tau$ is reflected with a ratio $r\sim1-e^{-\Delta/T}$ and the $\varepsilon$ is fully transmitted according to the interaction  {Eq.} (\ref{int}). By a Wick rotation $y\to it$, the collision matrices for Fig.  {\ref{fig1}} can be determined by the following 8-point Green's function
\begin{eqnarray}
&&\langle \tau(z_{1\tau})\varepsilon(z_{1\varepsilon})\tau(z_{2\tau})\varepsilon(z_{2\varepsilon}) \tau(z_{3\tau})\varepsilon(z_{3\varepsilon})\tau(z_{4\tau})\varepsilon(z_{4\varepsilon})\rangle\nonumber\\
&&\propto \langle \tau(z_{1\tau})\tau(z_{2\tau})\tau(z_{3\tau})\tau(z_{4\tau})\rangle\langle \varepsilon(z_{1\varepsilon})\varepsilon(z_{2\varepsilon})\varepsilon(z_{3\varepsilon})\varepsilon(z_{4\varepsilon})\rangle\nonumber 
\end{eqnarray}
with $z_{a\tau}\to  z_{a\varepsilon}$ for $z=x+iy$ because of the coset decomposition of SO(7)$_1$ CFT. The 4-point correlation function of a primary field $\phi$ in the CFT can be exactly calculated \cite{CFTbook}
\begin{eqnarray}
 \langle \phi(z_{1})\phi(z_{2})\phi(z_{3})\phi(z_4)\rangle=F(x)(z_{12}z_{13}z_{14}z_{23}z_{24}z_{34})^{-2h/3},
 \end{eqnarray}
 where $z_{ij}=z_i-z_j$ and $x=\frac{z_{12}z_{34}}{z_{13}z_{24}}$. The function $F(x)$ can be determined  {by} the Ward identities of the CFT. Exchanging any two $\phi$ fields, we have a statistical  phase $e^{2ih\pi}$ (where we have taken $e^{-i\pi}=-1$ for convenience). Wick rotating back, braiding $\tau$ or $\varepsilon$ obtains a statistical phase $e^{4i\pi/5}$ or $e^{i\pi/5}$.  In the following we assume $\tau$ is completely reflected while $\varepsilon$ is fully transmitted. In reality, the finite temperature will cause an error with probability proportional to $e^{-\Delta/T}$ in $\tau$ collision while the interaction deviating from (\ref{int}) can cause  {the} error both for $\tau$ and $\varepsilon$ collisions.}

\begin{figure}
\centerline{\includegraphics[width=0.5\textwidth]{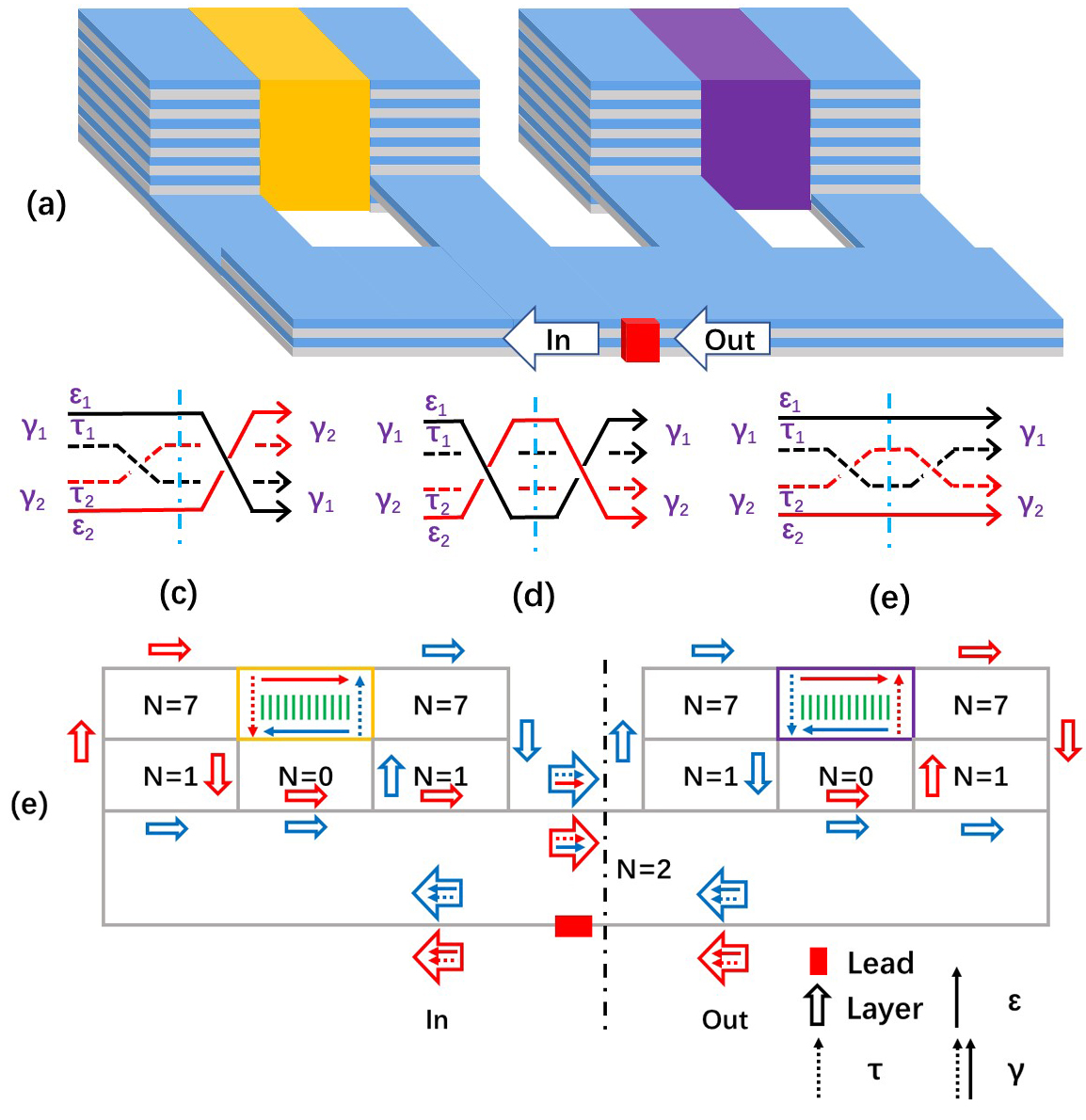}}
\caption{(Color online) The setup of the element $G(\theta)$. (a) The device setup with the orange and purple parts being the interacting domains. The blue {parts} are the TSC thin films while the grey {parts} are  the trivial insulator films. The red slab is the {lead} connected to external metal or to the other element in the TSC quantum circuit. (b), (c), and (d): The world lines between the in and out states, which represent $G(-\frac{\pi}{4})$, $G(\frac{\pi}{10})$ and $G(\frac{2\pi}{5})$, respectively. (e) The top view of the device for $G(-\frac{\pi}{4})$.\label{fig2}}
\end{figure}

 We now want to design the topological phase elements.  Fig. \ref{fig2} (a) is the schematics of the elements $G(\theta)=e^{-2i\theta}$. Fig. \ref{fig2} (e) is the top view of  Fig. \ref{fig2} (a) when the two interacting domains are different (The other two cases are shown in Appendix \ref{D}).
  A charged spinless fermion $\psi$ is injected from the {lead}  to the 2-layer $\chi$TSC {such} that $\psi\to \frac{1}{\sqrt2}(\gamma_1+i\gamma_2)$ where the $\chi$MEMs $\gamma_{1,2}$ run along the edges of layer 1 and 2, respectively. They then become the $R$ and $L$-$\chi$MEMs and enter one of the edge channels in 7-layers with an equal probability. The MEMs are factorized into anyons: $\gamma_{1,2}= \tau_{1,2}\varepsilon _{1,2}$.  In the yellow interacting domain, anyons, say $\tau$ from Fig. \ref{fig2} (e),  collide. This yields the braiding of $\tau$.  After the first collision, anyons run into the next 1-2-7 layer hybrid and collide in the purple domain. As a result, e.g., $\varepsilon$ anyons braid.
According to the anyon's conformal dimensions, the braiding of them causes statistical phases:  $\tau_1\to \tau_2, \tau_2\to e^{i\frac{4\pi}5}\tau_1$ and   $\varepsilon_1\to \varepsilon_2, \varepsilon_2\to e^{i\frac{\pi}5}\varepsilon_1$. The element $G(\theta)$ makes the anyons exchange twice: According to the world line in Fig. \ref{fig2} (b), exchanging $\tau$ and $\varepsilon$ in turn gives {$G(-\frac{\pi}4)$}, i.e.,
 \begin{eqnarray}
&& \tau_1\varepsilon_1+i \tau_2\varepsilon_2\to\tau_2\varepsilon_1+i e^{i\frac{4\pi}5} \tau_1\varepsilon_2  \to\tau_2\varepsilon_2 +i  e^{i\frac{4\pi}5} \tau_1e^{i\frac{\pi}5}\varepsilon_1\nonumber\\
&& =\tau_2\varepsilon_2-i \tau_1\varepsilon_1
=-i( \tau_1\varepsilon_1+i \tau_2\varepsilon_2 ).
 \end{eqnarray}
  This corresponds to $\gamma_1\to \gamma_2, \gamma_2\to-\gamma_1$ or $\gamma_{1,2}\to e^{-i\frac{\pi}2}\gamma_{1,2}$.
 Exchanging $\varepsilon$ twice gives $G(\frac{\pi}{10})$,     as shown in Fig. \ref{fig2}(c),
 \begin{eqnarray}
&& \tau_1\varepsilon_1+i \tau_2\varepsilon_2\to\tau_1\varepsilon_2+i \tau_2(e^{i\frac{\pi}5} \varepsilon_1)=\tau_1\varepsilon_1'+i\tau_2\varepsilon_2'\nonumber\\
&& \to \tau_1\varepsilon_2'+i\tau_2 e^{i\frac{\pi}5} \varepsilon_1'= e^{i\frac{\pi}5} (\tau_1\varepsilon_1+i \tau_2\varepsilon_2).
\end{eqnarray}
 This corresponds to $\gamma_a\to e^{i\frac{\pi}5} \gamma_a$. Exchanging $\tau$ twice gives $G(\frac{2\pi}{5})$ (Fig. \ref{fig2}(d)) and  corresponds to $\gamma_a\to e^{i\frac{4\pi}5} \gamma_a$.  Notice that $G(\frac{\pi}4)=(G(-\frac{\pi}4))^3$; $G(\frac{2\pi}5)=(G(\frac{\pi}{10}))^4$;  $G(-\frac{\pi}{10})=(G(\frac{\pi}{10}))^9$; and $G(-\frac{2\pi}5)=(G(\frac{\pi}{10}))^6$, and so on.\\

  \begin{figure}	
\centerline{\includegraphics[width=0.5\textwidth]{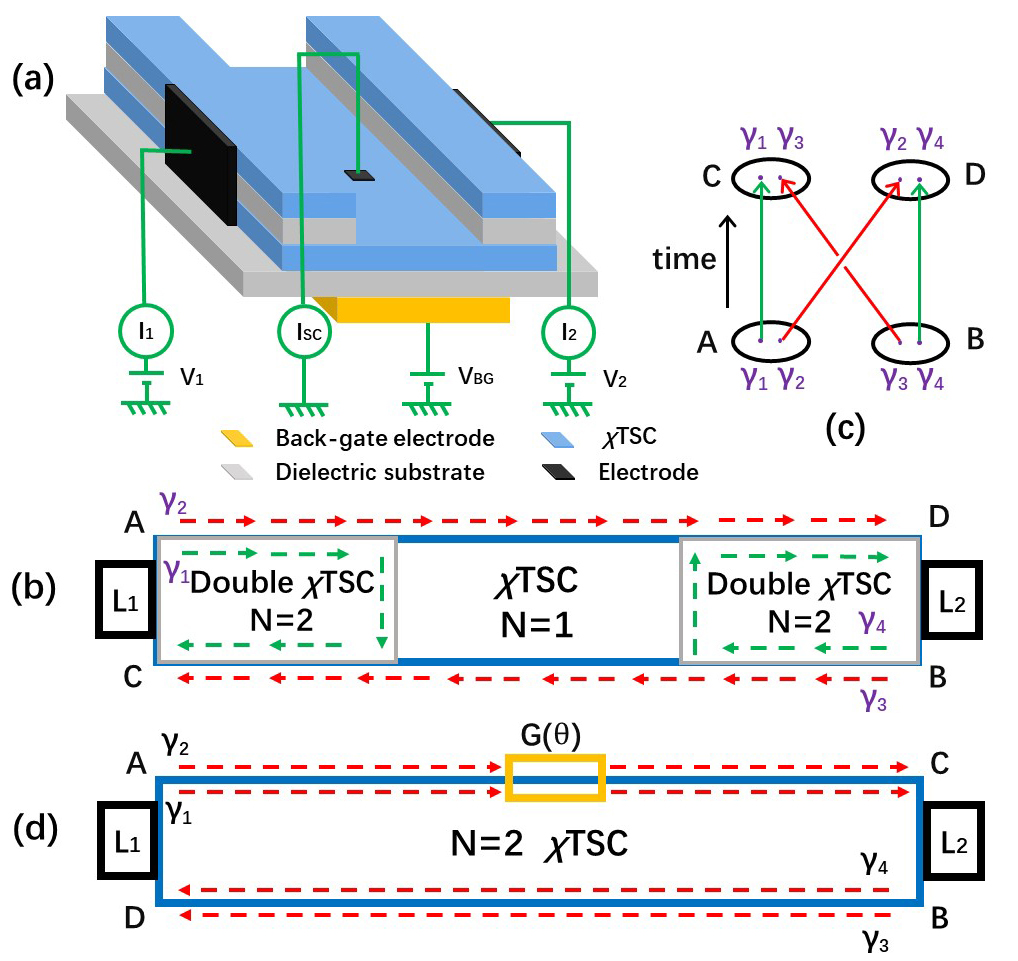}}
	\caption{(Color online) The device for the  basic braiding gates.
		(a) A schematic of a layered $\chi$TSC structure for the gate $B_{23}$. The black slabs are the {leads},  and the {blue and grey ones} are the same as those in the Fig. \ref{fig2}(a). The yellow {slab} is a back-gate electrode used to adjust the electric potential with a back-gate voltage when measuring the output data.  (b) The top view of the device in (a). The arrows stand for the {$\chi$MEMs}.  (c) The world line of the MEMs. (d) The device schematic of the phase gate. The yellow sticker  stands for the phase gate $G(\theta)$. \label{fig3}}
\end{figure}

\section{Universal set of topological quantum gates}

We now design a new set of topological quantum gates with these topological elements. Due to the FP conservation, the Hilbert space with a fixed FP for a single charged fermion is one-dimensional and then cannot encode a qubit. To construct a 1-qubit with a fixed FP, two charged fermion are required.

\subsection{The Basic Braiding gates and Hadamard Gate for 1-Qubit}

We consider charged fermions inject into the edges of two layer $\chi$TSC thin films from {leads} in the terminals  $A$ and $B$ (Fig. \ref{fig3}).   The initial states are then $|n_An_B\rangle$ for $n_{A,B}=0,1$ corresponding to the FP even and odd.  If we consider weak current limit, we assume $n_{A,B}=0,1$ are the fermion number and the state $|1_A\rangle=\psi_A|0\rangle=\frac{1}{\sqrt2}(\gamma_1+i\gamma_2)|0\rangle$ and $|1_B\rangle=\psi_B|0\rangle=\frac{1}{\sqrt2}(\gamma_3+i\gamma_4)|0\rangle$. The basis of the initial state space is then
given by $$\{|0_A^{\gamma_1\gamma_2}0_B^{\gamma_3\gamma_4}\rangle ,|0_A^{\gamma_1\gamma_2}1_B^{\gamma_3\gamma_4}\rangle ,|1_A^{\gamma_1\gamma_2}0_B^{\gamma_3\gamma_4}\rangle,|1_A^{\gamma_1\gamma_2}1_B^{\gamma_3\gamma_4}\rangle\}.$$
The first and fourth ones form 1-qubit with the {FP} even while the other 1-qubit has the FP odd. We now braid two of MEMS while keeping the other two run straightforwardly. The basis of the final state spaces are transformed as
$$\{|0_C^{\gamma_2\gamma_1}0_D^{\gamma_3\gamma_4}\rangle ,|0_C^{\gamma_2\gamma_1}1_D^{\gamma_3\gamma_4}\rangle ,|1_C^{\gamma_2\gamma_1}0_D^{\gamma_3\gamma_4}\rangle,|1_C^{\gamma_2\gamma_1}1_D^{\gamma_3\gamma_4}\rangle  \},$$
$$\{|0_C^{\gamma_1\gamma_3}0_D^{\gamma_2\gamma_4}\rangle ,|0_C^{\gamma_1\gamma_3}1_D^{\gamma_2\gamma_4}\rangle ,|1_C^{\gamma_1\gamma_3}0_D^{\gamma_2\gamma_4}\rangle,|1_C^{\gamma_1\gamma_3}1_D^{\gamma_2\gamma_4}\rangle  \},$$
$$\{|0_C^{\gamma_1\gamma_2}0_D^{\gamma_4\gamma_3}\rangle ,|0_C^{\gamma_1\gamma_2}1_D^{\gamma_4\gamma_3}\rangle ,|1_C^{\gamma_1\gamma_2}0_D^{\gamma_4\gamma_3}\rangle,|1_C^{\gamma_1\gamma_2}1_D^{\gamma_4\gamma_3}\rangle  \},$$
which correspond to braiding $(\gamma_1,\gamma_2),(\gamma_2,\gamma_3),$ and $(\gamma_3,\gamma_4)$, respectively.
For a fixed FP, there are braiding matrices which transform the initial states to the final states, e.g., the braiding matrix for braiding $\gamma_2$ and $\gamma_3$ for the FP odd state is given by
\begin{equation}
		\left(
		\begin{array}{ccc}
		|0_C^{\gamma_1\gamma_3}1_D^{\gamma_2\gamma_4}\rangle \\
		|1_C^{\gamma_1\gamma_3}0_D^{\gamma_2\gamma_4}\rangle
		\end{array}
		\right)=B^{(-)}_{23}\left(
		\begin{array}{ccc}
		|0_A^{\gamma_1\gamma_2}1_B^{\gamma_3\gamma_4}\rangle \\
		|1_A^{\gamma_1\gamma_2}0_B^{\gamma_3\gamma_4}\rangle
		\end{array}
		\right).
		\end{equation}
And also we can define all $B^{(\pm)}_{ab}$ for $a<b=1,...,4$, where $\pm$ stands for the FP of the {qubit}.
 Lian et al. gave a proposal for the gate $B^{(-)}_{23}$ with a quantum anomalous Hall insulator/superconductor proximity structure \cite{BL}. We here use the layered $\chi$TSC structure for $B^{(-)}_{23}$ (Fig. \ref{fig3}(a)).  As shown in Fig. \ref{fig3}(b),  in the edges of the upper layer, $\gamma_1$ runs from $A$ to $C$ while $\gamma_4$ from $B$ to $D$ while  in the edges of the lower layer, $\gamma_2$ runs from $A$ to $D$ while $\gamma_3$ from $B$ to $C$. This yields the braiding between $\gamma_2$ and $\gamma_3$ (Fig. \ref{fig3}(c)).
 Under a braiding operation $\gamma_2\rightarrow\gamma_3, \gamma_3\rightarrow-\gamma_2$, the evolution of the FP odd state is thus equivalent to a $B_{23}^{(-)}$ gate ({Fig.~\ref{fig3}(c)})
		\begin{equation}
		B_{23}^{(-)}=\frac{1}{\sqrt{2}}\left(
		\begin{array}{ccc}
		i  & 1 \\
		1 &  i
		\end{array}
		\right). \label{B23}
		\end{equation}
One can also show that  $B_{23}^{(+)}=B_{23}^{(-)}\equiv B_{23}$ which is FP-independent. Exchanging $\gamma_1$ and $\gamma_2$ can be simply realized by adding a phase gate on the upper edge of the double layer thin films in Fig. \ref{fig3} (d) with $G(\theta)=G(-\frac{\pi}4)$. The corresponding phase gates $B_{12}^{(\pm)}$ are also the same, i.e., $B_{12}={\rm diag}(1,i)$, which is the $\frac{\pi}4$-phase gate.  The braiding matrix $B_{34}$, which is realized by adding  $G(-\frac{\pi}4)$ on the lower edge instead of adding it on the upper edge in Fig. \ref{fig3} (d). However, we find that $B_{34}^{(-)}$ is not the same as $B_{34}^{(+)}$ because $B_{34}^{(+)}={\rm diag}(1,i)$ while $B_{34}^{(-)}={\rm diag}(i,1)$.

With these basic braiding {matrices}, we can make the Hadamard gate, up to a global phase $i$,
\begin{eqnarray}\label{H}
B_{12}B_{23}B_{12}\sim H=\frac{1}{\sqrt2}\left(
		\begin{array}{ccc}
		1 & 1 \\
		1 & -1
		\end{array}
		\right).
	\end{eqnarray}		
And also  $H\sim B^{(-)}_{34}B_{23}B^{(-)}_{34}\sim B^{(+)3}_{34}B_{23}B^{(+)3}_{34} $ up to a sign. Therefore, the Hadamard gate is also FP-independent when the basis for {1-qubit} is properly chosen.  Furthermore, we have $Z=B_{12}^2$, $X=HZH$ and $Y=(ZH)^2$. Up to a global phase, they all are independent of the FP. In  this way, we have a set of FP-independent Clifford gates.

In general, Fig. \ref{fig3} (d) gives a phase gate $B(\theta/2)={\rm diag}(1, G(-\theta))\sim{\rm diag}(G(\theta/2),G(-\theta/2))$. The topological phase gates through anyon braiding are $B(-\frac{\pi}{10})$, $B(-\frac{2\pi}5)$ as well as $B(\frac{\pi}{10})=B^9(-\frac{\pi}{10})=B^{-1}(-\frac{\pi}{10})$,  and so on. They are independent of the FP of the 1-qubit.

 \begin{figure}
\centerline{\includegraphics[width=0.55\textwidth]{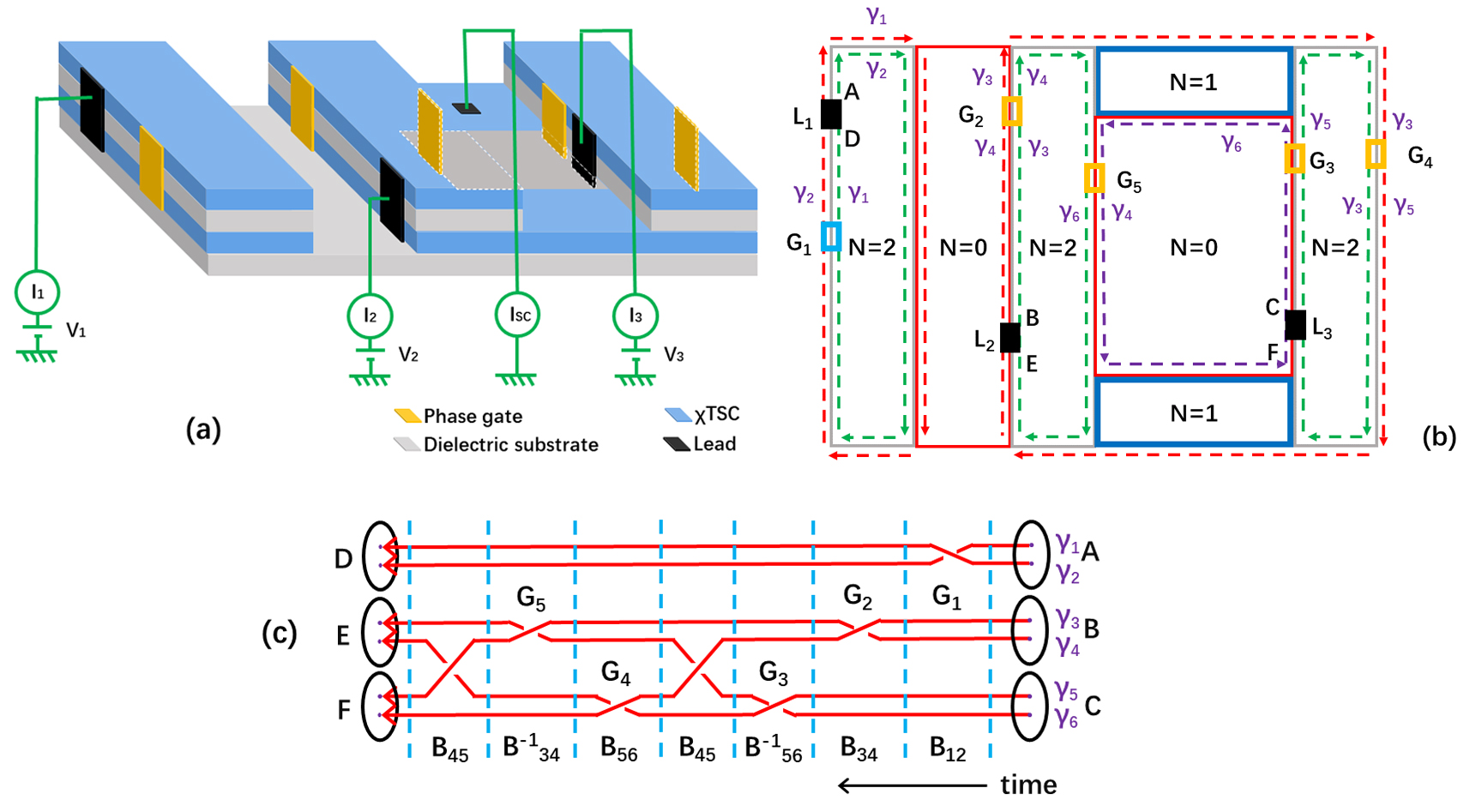}}
		\caption{(Color online)
		The device for the CNOT.  (a) A schematic of a layered $\chi$TSC structure for the CNOT gate. The electric potential of the  $\chi$TSC has been set to zero. (b) The top view of (a). $G_a, (a=1,..,5)$ are 5 phase elements. Yellow stickers represent $G(-\frac{\pi}4)$, and blue one is $G(\frac{\pi}4)$. (c) The world line of the CNOT.\label{fig4} }
\end{figure}

\subsection{The CNOT Gate}	

To achieve universal TQC, we design a CNOT gate for 2-qubits using six $\chi$MEMs $\gamma_1,...,\gamma_6$. The setup of the CNOT gate is FP-dependent.  We first take the basis with even total FP as an example, i.e., the incoming basis
\begin{eqnarray}
(|0_A0_B0_C\rangle, |0_A1_B1_C\rangle, |1_A0_B1_C\rangle, |1_A1_B0_C\rangle)^T \label{incoming}
\end{eqnarray}
 and the outgoing basis
\begin{eqnarray}
(|0_D0_E0_F\rangle, |0_D1_E1_F\rangle, |1_D0_E1_F\rangle, |1_D1_E0_F\rangle)^T. \label{out}
\end{eqnarray}
The device is shown in Fig.~\ref{fig4} (a) and its top view in Fig.~\ref{fig4}(b).
If we use the elements $G_a(\theta_a)$ with $\theta_{1}=\frac{\pi}4$ and $\theta_{2,3,4,5}=-\frac{\pi}4$, the braiding matrices for 2-qubits are $B_{12}^{(2)}={\rm diag}(1,1,i,i)$, $B_{34}^{(2)}={\rm diag}(1,i,1,i)$, and  $B_{56}^{(2)}={\rm diag}(1,i,i,1)$ (See Fig.~\ref{fig4}(c)). $B_{45}^{(2)}$ in Fig.~\ref{fig4}(c) is the 2-qubits counterpart of the $B_{23}$ gate,
\begin{eqnarray}
		B_{45}^{(2)}=
		\frac{1}{\sqrt{2}}\left(
		\begin{array}{cccc}
		i  & 1 & 0 & 0 \\
		1  & i & 0 & 0 \\
		0  & 0 & i & 1 \\
		0 & 0 & 1 & i
		\end{array}
		\right) \nonumber
		\end{eqnarray}
where the subscripts stand for the relative positions of the {$\gamma$s} at a given time slice. According to the sequence of the $B$-matrices in Fig.~\ref{fig4}(c), we obtain the CNOT gate
\begin{eqnarray}
{\rm CNOT}^{(+)}&=&B_{45}^{(2)}B^{(2)}_{34}B^{(2)}_{56}B^{(2)}_{45}B^{(2)}_{56}B^{(2)}_{34}B^{(2)-1}_{12}
		\nonumber\\ &=&\left(
		\begin{array}{cccc}
		1  & 0 & 0 & 0 \\
		0  & 1 & 0 & 0 \\
		0  & 0 & 0 & 1 \\
		0  & 0 & 1 & 0
		\end{array}
		\right). ~~\label{cnot+}
		\end{eqnarray}
When the input state is chosen as the odd FP 2-qubits, the device in Fig. \ref{fig4} does not produce the CNOT gate but  the matrix
\begin{equation}	
 i\left(\begin{array}{cccc}
		0& 1 & 0 & 0 \\
		1& 0 & 0 & 0 \\
		0  & 0 & -1 & 0 \\
		0  & 0 & 0 & -1
		\end{array}
		\right).
 \end{equation}
The controlling qubit and the target qubit exchange while a sign difference exists. This cannot fulfill the logical task as the CNOT gate. If we input the odd FP 2-qubits, to have the CNOT$^{(-)}$, the phase gates should be reassigned: We use the elements $G_{1,2,4}(\theta_1)=-\frac{\pi}{4}$ and $G_{3,5}(\theta_{3,5})=\frac{\pi}{4}$. This gives
\begin{eqnarray}\label{cnot-}
B_{45}^{(2)}B^{(2)-1}_{34}B^{(2)}_{56}B^{(2)}_{45}B^{(2)-1}_{56}B^{(2)}_{34}B^{(2)}_{12}=\left(
		\begin{array}{cccc}
		1  & 0 & 0 & 0 \\
		0  & 1 & 0 & 0 \\
		0  & 0 & 0 & 1 \\
		0  & 0 & 1 & 0
		\end{array}
		\right).~~
\end{eqnarray}

\subsection{Universal Topological Quantum Gates}\label{universal}

The authors of Ref. \cite{vatan} proved the universality of the quantum circuit models  {which is equivalent} to Ref. \cite{UTQC}.
They chose three gates $\{H, Z^{\frac{1}{4}}=B(\frac{\pi}8), {\rm CNOT}\}$ as their universal set. In their proof, the key essence is that they can create two orthogonal axes and a phase with irrational number times $\pi$ from the universal set, namely, any element in SU(2) can be approximated in a desired precision. Here we will show the existence of the orthogonal axes and the irrational number for strongly correlated Majorana fermion based the TQC.

The universal set that we choose is $\{H,\sqrt{Z}= B(\frac{\pi}{4}), B(-\frac{\pi}{10}),{\rm CNOT}\}$.  They all have been obtained in last two subsections. Therefore, our universal set is topological and fault-tolerant. Similar to those in \cite{vatan},  we use
\begin{equation}
 X^{\frac{2}n}=HB(\frac{\pi}n) H,
\end{equation}
and define
\begin{eqnarray}
F_{2/5}=X^{2/5}B(\frac{\pi}4)X^{-2/5},
\end{eqnarray}
and
\begin{widetext}
\begin{eqnarray}
U_1=e^{i\lambda\pi\hat{n}_1\cdot\vec{X}}&\equiv& B(-\frac\pi{10})X^{\frac{1}{5}}=\frac{1}{8}\left(
\begin{array}{cc}
5+\sqrt{5}+i \sqrt{10-2 \sqrt{5}} & 3-\sqrt{5}-i \sqrt{10-2 \sqrt{5}} \\
-3+\sqrt{5}-i \sqrt{10-2 \sqrt{5}} & 5+\sqrt{5}-i \sqrt{10-2 \sqrt{5}} \\
\end{array}
\right) \label{U1}
,\\
U_2=e^{i \lambda\pi\hat{n}_2\cdot\vec{X}}&\equiv& F_{2/5}^{-1}U_1F_{2/5}=\frac{1}{32}
\left(
\begin{array}{cc}
(4+4 i) \sqrt{5}+20 & \left(\sqrt{5}-1\right) \left(8 i+\sqrt{10-2 \sqrt{5}}\right) \\
-\left(\sqrt{5}-1\right) \left(-8 i+\sqrt{10-2 \sqrt{5}}\right) & (4-4 i) \sqrt{5}+20 \\
\end{array}
\right), \  \label{U2}
\end{eqnarray}
\end{widetext}
where $\vec X=(X,Y,Z)$; $\lambda_i$ and unit vectors $\hat{n}_i$ are to be determined. 
Thus, solving {Eqs.} (\ref{U1}) and (\ref{U2}), we find that $\lambda_1=\lambda_2=\lambda$ where $\lambda$ is given by
\begin{eqnarray}\cos(\lambda\pi)=\cos^2\frac{\pi}{10}=\frac{5+\sqrt{5}}{8},
\end{eqnarray}
while
\begin{eqnarray}
&&\hat{n}_1\approx(-0.293893, 0.0954915, 0.293893), \nonumber\\
&&\hat{n}_2\approx(0.309017, 0.0908178, 0.279508).\nonumber
\end{eqnarray}
Clearly, $\hat{n}_1\cdot\hat{n}_2=0$. Furthermore, $e^{i2\lambda\pi}$ is the non-cyclotomic root of the irreducible monic polynomial equation $x^4+x^3/4+x^2/16+x/4+1=0$. Hence, $\lambda$ is irrational. If constructing a local isomorphism from SU(2)  to SO(3), then any element of SU(2)  can be expressed as an Euler rotation. That is, if $\alpha,\beta$ and $\gamma$ are the Euler angles, any elements of SU(2) can be written as \cite{Euler}
$
{e}^{i\phi\hat{n}\cdot\vec X }={e}^{i\alpha {\hat n}_1\cdot {\vec X}}{e}^{i\beta {\hat n}_2\cdot\vec X }{e}^{i\gamma {\hat n}_1\cdot \vec X}.
$
Because $\lambda$ is irrational, there are $a,b$ and $c$ {such} that $\alpha\approx a\lambda\pi,\beta\approx b\lambda\pi$ and $\gamma\approx c\lambda\pi$, i.e., 
\begin{eqnarray}\label{Euler}
 {e}^{i\phi\hat{n}\cdot\vec X }\approx ({{U}_{1}})^{a}({{U}_{2}})^{b}({{U}_{1}})^{c},
\end{eqnarray}
where $a,b$ and $c$ are integers determined by $\alpha,\beta,$ and $\gamma$ in a desired precision. This gives the proof of the universality of $\{H,\sqrt{Z}= B(\frac{\pi}{4}), B(-\frac{\pi}{10}),{\rm CNOT}\}$ according to \cite{UTQC}.

\section{Electric signals of the  outputs }
Since the initial and final states are the charged fermions, the inputs and readout of the designed TQC are electric.  To read out the computation results of the TQC, we must translate the outgoing states of the quantum gate operations into electric signals. 

 For the $ZH$ gate, the conductance between the leads at the two ends of the device measures the operating result \cite{BL}. For the other 1-qubit gates, the conductance calculation is also direct.  Here, we give an example for the 2-qubits, the CNOT gate. The incoming and outgoing bases are given by (\ref{incoming}) and  (\ref{out}), respectively.
The FP $0$ or $1$ can be read out by the electric signals at the leads. The CNOT gate changes $|1_A0_B1_C\rangle$ to $|1_A1_B0_C\rangle$ and vise versa, while keeping $|0_A0_B0_C\rangle$ and $|0_A1_B1_C\rangle$ unchanged.
Thus, these states changes can be read out from the conductance between {Lead 2 and Lead 3}: $\sigma_{23}=(1-\langle \psi_{out}|\psi_{in}\rangle)\frac{e^2}h$. Namely, $\sigma_{23}=e^2/h$ for $|\psi_{in}\rangle=|1_A0_B1_C\rangle$ or $|1_A1_B0_C\rangle$  while $\sigma_{23}=0$ for $|\psi_{in}\rangle= |0_A0_B0_C\rangle$ or $|0_A1_B1_C\rangle$.

 If the phases in $G_a(\theta_a)$  in Fig. \ref{fig4} are arbitrary,   the  outgoing state $|\psi_{out}\rangle$ is given by $|\psi_{out}\rangle=U(\theta_a)|\psi_{in}\rangle$ where $|\psi_{in}\rangle $ is the incoming  state and  $U(\theta_a)$ is the unitary transformation (See  Appendix \ref{F}). For example,  for an incoming state $|\psi_{in}\rangle=|0_A0_B0_C\rangle$, the outgoing state is
$
|\psi_f\rangle=-\frac{1}{2}(1-e^{-i \text{$\theta_{5}$}})|0_A0_B0_C\rangle+\frac{1}{2}[ie^{-i \text{$\theta_{12}$}} (1+e^{-i \text{$\theta_{5}$}})]|0_A1_B1_C\rangle
$
and the corresponding conductance is
$
\sigma_{23}=\cos^2(\theta_5)\frac{e^2}{h}.$ For the topological CNOT gate, these $G_a(\theta_a)$ are given by the device in  $\theta_2=\theta_3=\theta_4=\theta_5=-\frac{\pi}{4}$ and $\theta_1=\frac{\pi}{4}$, and  $\sigma_{23}$ exactly gives the result we analyzed before.

 Before ending this section,  {we would} like to discuss the spin polarization problem. We assume the $p_x+ip_y$ $\chi$TSC is a spinless fermion system. In reality, it should be a spin polarized system. First we would like to emphasize that a calculating unit, i.e., the quantum circuit, must be   {in connection to one bulk of TSC} in order to prevent the phase loss of the phase gates.
 Therefore, the spin of the Cooper pairs in the circuit is all of the same polarization and so the spin of the chiral edge states is.  A real spin polarized $\chi$TSC material is not yet ready in nature. If it was found, the Cooper pair's spin would be fixed by the materials. The electron's spin from Leads to the edge of TSC will be automatically selected by the TSC. A better choice is using spin selecting valve, e.g., injecting in and putting out the electrons from a ferromagnet \cite{ferro}, anomalous Rashba metal/superconductor junctions \cite{ARM/SC} or in noncollinear antiferromagnets \cite{antiferro},
 in order to prevent the opposing spin electrons affect the data read in and out.  In these cases, a gap opens between the spin majority electron and spin minority electron near the Fermi level. The conversion losses mainly depend on the mismatch of Fermi velocities, the polarizability, and the energy of electronics. Lower mismatch of Fermi velocities and higher polarizability are expected to get fewer losses.

 \section{The quantum circuits with topological quantum gates}

\begin{figure}[!h]
        \centering \includegraphics[width=0.9\columnwidth]{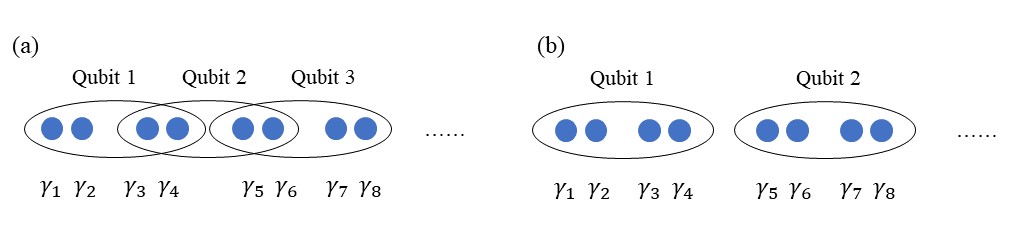}
        \caption{(Color online) Two different encodings \cite{DSFN}.(a) The dense encoding process using $2n+2$ MEMs to encode $n$-qubits;(b) The sparse encoding, $n$-qubits are encoded by $4n$ MEMs.
        }\label{figsd}
\end{figure}

 We have constructed a set of universal quantum gates for 1-qubit and the CNOT gate. In principle, we can use them to construct the quantum circuit models associated with quantum algorithms, e.g., the quantum Fourier transformation, a classical adder with quantum gates, and then Shor's integer factorization algorithm. However, the conventional TQC process with non-Abelian anyon does not follow the quantum circuit models. In the former, a larger qubit gate is not simply constructed by the smaller gates. Our CNOT gate construction is an example because it is not assembled by two 1-qubit gates. In fact,  an $n$-gate in this quantum computation uses $2n+2$ MEMs (Fig. \ref{figsd} (a)). This is called the `dense encoding' process \cite{DSFN}. With the quantum circuit model, an $n$-gate is constructed by $4n$ MEMs (Fig. \ref{figsd} (b)).  This process  matches with the quantum circuit model and is called the `sparse encoding' process \cite{DSFN}. In order to encode Shor's algorithm, we take the sparse encoding. Since the set of universal quantum gates for the 1-qubit is not dependent on the FP of the 1-qubit, they are good elements for the quantum circuit. However, in the sparse encoding, one can only prepare qubits in superposition states, but no entanglement states with braiding only \cite{Bravyi}.  For instance, we cannot make the CNOT gate with two gates for 1-qubit. It has to be realized in the dense encoding as we have shown. Therefore, our process in fact is a mixed one of the sparse-dense encoding. We use six MEMs to construct a CNOT gate. This does not match the quantum circuit model. On the other hand, the CNOT gate we constructed in Sec. V B is dependent on the FP of the 2-qubits. To measure the FP of the 2-qubits, two additional ancillary MEMs are required. This makes a CNOT gate with eight MEMs in the sparse encoding. Adding this sparse encoding CNOT gate to the universal gates for 1-qubit, we can make a quantum circuit as usual.

To illustrate such a process, we show the quantum circuit with our TQC process for Shor's integer factorization algorithm.

\subsection{The CNOT Gate in Sparse Encoding}

We construct the CNOT gate in the sparse encoding. As we mentioned, two ancillary MEMs will be introduced to measure the FP of the {2-qubits}. Notice that the measurements here do not measure the fermion occupation number of the computational states directly, but measure the parity operator $ \hat{P}=i{\gamma}_{1}{\gamma}_{2}... $ which won't destroy the coherence of the quantum state. In experiment, we make a side measurement to the total charge or the total spin for the subsystem to determine the eigenvalue of $\hat P$ \cite{P measure}.

\begin{figure}[!h]
        \centering \includegraphics[width=1\columnwidth]{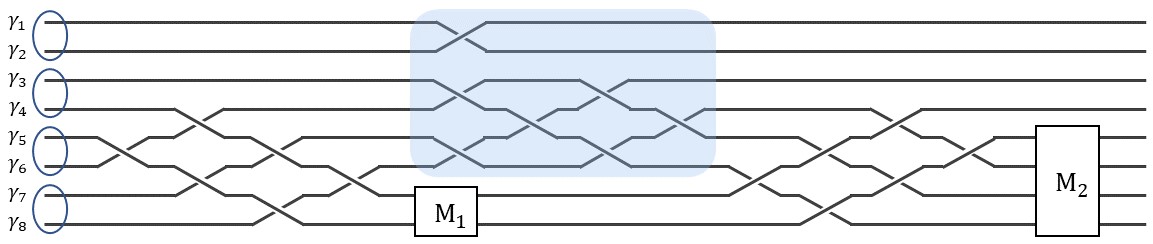}
        \caption{(Color online) The CNOT in sparse encoding with even parity. The blue shade area is the same as the dense encoding CNOT$^{(+)}$ (Fig. \ref{fig4} (c)). Because of the no-entanglement rule, two FP measurements are needed to get the entanglement states.
        }\label{S-CNOT1}
\end{figure}

More specifically, we want to measure the FP of 2-qubits that are input into a CNOT$^{(+)}$ given by Eq. (\ref{cnot+}).  In the sparse encoding, 2-qubits are associated with 4 pairs MEMs $ {\gamma}_{1},{\gamma}_{2},...,{\gamma}_{8} $, we use the same process taken in Ref. \cite{DSFN}, i.e.,  we enforce ${\gamma}_{1}{\gamma}_{2}{\gamma}_{3}{\gamma}_{4}={\gamma}_{5}{\gamma}_{6}{\gamma}_{7}{\gamma}_{8}=+1 $ (or $-1$). We take $\gamma_{4,5}$ as an ancillary qubit and  measure its FP, i.e., $i{\gamma}_{4}{\gamma}_{5}$ (Fig. \ref{S-CNOT1}). Thus, $\gamma_{6}$ exchanges with $\gamma_{4,5}$ and forms a new pair with $\gamma_{3}$.

If $i{\gamma}_{4}{\gamma}_{5}= 1$ (or $-1$), the remaining six MEMs form  2-qubits in the dense encoding with even FP. We then use the 2-qubits as the input state to a CNOT$^{(+)}$ given by Eq. (\ref{cnot+}),  according to the selected parity of the CNOT$^{(+)}$ (See Fig. \ref{S-CNOT1}).  Otherwise, we do the measurement again until $i{\gamma}_{4}{\gamma}_{5}= 1$ (or $-1$).  Finally, we add ${\gamma}_{4,5}$ back and exchange $\gamma_{6}$ with $\gamma_{4,5}$ again, then measure ${\gamma}_{5}{\gamma}_{6}{\gamma}_{7}{\gamma}_{8}$. If it is $1$ (or $-1$), the eight MEMs return to a 2-qubits with the positive FP in the sparse encoding  which is the output state (Fig. \ref{S-CNOT1}).  In this way, we encode the CNOT gate with the sparse encoding. For more details, see Appendix \ref{CNOTdetial}.

We can improve the efficiency in inputing. After making the measurement $M_1$, we can switch $\theta_1$ from $-\frac{\pi}4$ to $\frac{\pi}4$ and $\theta_{3,5}$ from $-\frac{\pi}4$ to $\frac{\pi}4$ if $i\gamma_4\gamma_5=-1 ({\rm or}~ 1)$  as we will see in Sec. VIII. In this case, the CNOT$^{(+)}$ gate becomes the CNOT$^{(-)}$.  Thus, instead of abandoning the FP negative state, it can be used. We next add back  $\gamma_{4,5}$ and do the similar operations to return the 2-qubits in the sparse encoding as before. 
Furthermore, we can further improve the efficiency if the FP measurement in $M_2$ is wrong by correcting FP, which will not be studied here. 
  
\subsection{Brief Introduction of Shor's Algorithm}

In the coming several subsections, we describe Shor's integer factorization algorithm with our TQC process. We first give a brief introduction to Shor's algorithm \cite{Nelson}.

The key step of Shor's algorithm is turning the integer factorization problem to a period-finding problem and then uses quantum computer to realize a period-finding subroutine, i.e., to find the period $r$ of the function $f(x)=a^{x}{\rm mod} N$, where $N$ is the integer waiting to be factorized and $a$ is an arbitrary number that is coprime with $N$. Calculating the greatest common divisors $n_1={\rm gcd}(a^{\frac{r}{2}+1},N)$ and $n_2={\rm gcd}(a^{\frac{r}{2}-1},N)$, then $N={n_1}\cdot {n_2}$, which can be realized by Euclid's algorithm classically.

We denote the basis of a 1-qubit with a given FP, e.g.,  $\{|0^{\gamma_1\gamma_2}1^{\gamma_3\gamma_4}\rangle ,|1^{\gamma_1\gamma_2}0^{\gamma_3\gamma_4}\rangle\}$ as $\{|0\rangle,|1\rangle\}$. Starting from two registers in which Register 1 $=|0^l\rangle$ which is the product of $l$  $|0\rangle$ and Register 2$=|0^q\rangle$. The quantum part of Shor's algorithm, the period-finding subroutine, is composed by three steps \cite{Nelson}:
(i) Doing the quantum Fourier transformation to Register 1 with $2^q=N$, $n=2^l$ is a number {within} $(N^2,2N^2)$ 
\begin{eqnarray}
|0^l\rangle|0^q\rangle\to (F_n|0^l\rangle)|0^q\rangle=\frac1{\sqrt n}\sum_{x=0}^{n-1}|x\rangle|0^q\rangle
\end{eqnarray}
where $|x\rangle$, e.g., $|0\rangle=|0...0\rangle, |1\rangle=|0...01\rangle, |2\rangle=|0...011\rangle$, etc.,  label the basis vectors of $l$-qubits. That is, the basis vector $|x\rangle$ indicates a binary number corresponding to $x$. 

(ii) Doing the modular exponentiation $U_f$ {such} that
 \begin{eqnarray}
U_f\frac1{\sqrt n}\sum_{x=0}^{n-1}|x\rangle|0^q\rangle=\frac1{\sqrt n}\sum_{x=0}^{n-1}|x\rangle|f(x)\rangle
\end{eqnarray}
Making a measurement on Register 2 {such} that  the state in Register 1 collapses to $$\frac{1}{\sqrt m}\sum_{j=0}^{m-1}|jr+s\rangle$$
 where $s$ is some number less than $r$ and $m$ is the number of $f(x)=f(s)$ with $x\in \{0,...,n-1\}$.

 (iii) Doing the quantum Fourier transformation again to Register 1 and measuring $$F_n\frac{1}{\sqrt m}\sum_{j=0}^{m-1}|jr+s\rangle,$$  we obtain a random number $y$. Repeat the process again, we have a new $y$. The maximal one in a series of $y$ is the period $r$.

\begin{figure}[!h]
        \centering \includegraphics[width=0.7\columnwidth]{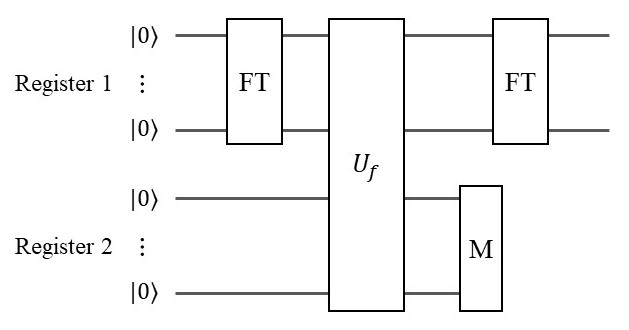}
        \caption{The period-finding subroutine. It is composed by three steps: Do the quantum Fourier transformation (FT) to Register 1; apply the modular exponentiation $U_f$ and measure Register 2 {such} that the state in Register 1 collapses.  Do the quantum Fourier {transformation} to Register 1 again. Finally, measure the output state in Register 1.
        }\label{Shor}
\end{figure}

 Fig. \ref{Shor} is the schematic of Shor's period-finding algorithm.  In the following we explain the detailed algorithms for the quantum Fourier transformation and the modular exponentiation with our quantum circuit model.

\subsection{Quantum Fourier Transformation}

\begin{figure}[!h]
        \centering \includegraphics[width=1\columnwidth]{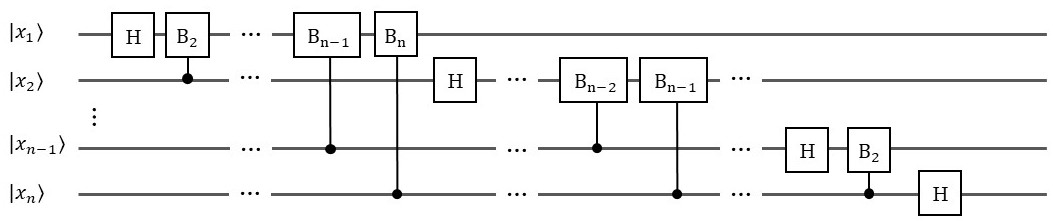}
        \caption{The circuit implementation for quantum Fourier {transformation}.
        }\label{QFT}
\end{figure}

Quantum Fourier {transformation} is the quantum analogue of the discrete Fourier transformation, which plays an important role in many quantum algorithms \cite{Nelson}. The circuit implementation is shown in Fig. \ref{QFT}. It consists of the Hadamard gate and the controlled-phase gate $controlled-B_n$ with $B_n=B(-\frac{\pi}{2^n})$.

The controlled-phase gate is a two-qubit gate, i.e., the target qubit is acted by a phase gate,
\begin{equation} \label{CG1} {\rm controlled}-B_n=
\left(
\begin{matrix}
1 & 0 & 0 & 0 \\
0 & 1 & 0 & 0 \\
0 & 0 & {e}^{-i\pi/{{2}^{n}}} & 0 \\
0 & 0 & 0 & {e}^{i\pi/{{2}^{n}}}
\end{matrix}
\right)
\end{equation}

\begin{figure}[!h]
        \centering \includegraphics[width=0.7\columnwidth]{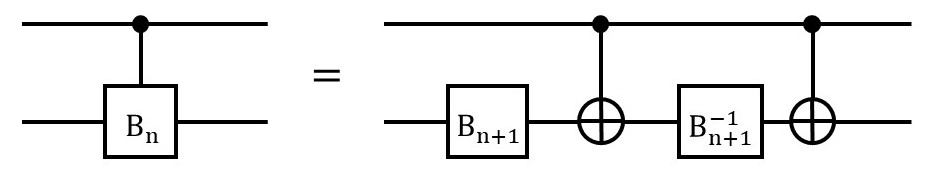}
        \caption{ The decomposition of the controlled-$B_n$.
        }\label{C-G1}
\end{figure}

Controlled-gates are a kind of gates often used in quantum computation and are proven to be decomposed into the combination of CNOT and 1-qubit gates \cite{UTQC}. In case of the controlled-$B_n$, one of the decomposition methods is given by Fig. \ref{C-G1}, i.e., the controlled-${B_n}$ is composed by the 1-qubit gate $B_{n+1}$,
\begin{equation}
B_{n+1}=
\left(
\begin{matrix}
{e}^{-i\pi/{{2}^{n+1}}} & 0 \\
0 & {e}^{i\pi/{{2}^{n+1}}}
\end{matrix}
\right)
\end{equation}

We see that $B_{n+1}$ is a group element of SU(2).  For example, for the controlled-$B_2$ gate ($B_2$ is often called the $S$ gate or $\frac{\pi}4$-gate), $B(\frac{\pi}{8})$ (also called the $\frac{\pi}8$-phase gate  or  $T$-gate) is needed. As we have shown, it can be approximated by Eq. (\ref{Euler}). When $a=18,b=22$ and $c=17$, Eq. (\ref{Euler}) gives the $T$-gate in a precision $>97\%$.

Any 1-qubit quantum gate can be approximated by Eq. (\ref{Euler}) and a rough calculation can estimate the circuit scale.
Suppose a quantum gate within the accuracy range $\sigma$ is needed. This is equivalent to move a vector ${\bf A}$ on a Bloch sphere to a designated point by rotating about two orthogonal axes $M=a+b+c$ times successively. $M$ operations will give a total of $\frac{1}{6}(M+1)(M+2)(M+3)$ different points on the Bloch sphere. It requires that these points can cover the sphere. Defining the accuracy requirement as the solid angle range of the vector neighborhood, then  $\frac{1}{6}(M+1)(M+2)(M+3)\cdot\sigma\sim4\pi $. With the accuracy requirement $\sigma$, it takes about $M\sim\sqrt[3]{\frac{24\pi}{\sigma}} $ steps to get the desired quantum gate. This argument, however, doesn't depend on the specific gate, but only related to the accuracy, is an estimate of the upper limit. For some special gate, much less operations are required.

\subsection{Adder and Toffoli Gate}

\begin{figure}[!h]
        \centering \includegraphics[width=0.5\columnwidth]{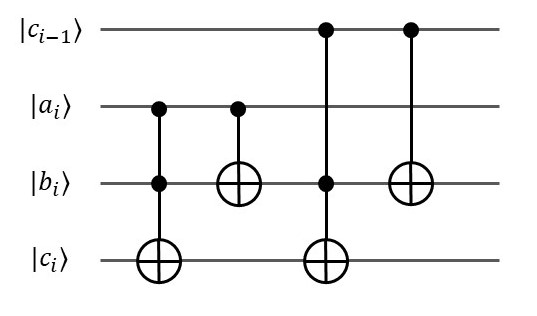}
        \caption{The construction of adder. $|a_i\rangle$ and $|b_i\rangle$ are the $i$th qubit of Register 1 and Register 2, respectively. $|c_i\rangle$ represents the $i$th qubit of the ancillary register to carry the calculation.
        }\label{adder}
\end{figure}

As we have seen, the modular exponentiation applies following {transformation}
$$ U_f|x\rangle\otimes|0^q\rangle\to |x\rangle\otimes|a^x{\rm mod}N\rangle, $$
where $|x\rangle$ and $|0^q\rangle$ represent two registers and the $U_f$ is determined by $a$ and $N$.
The elementary block for a computer to realize basic arithmetic operations is the addition unit or called an adder. Starting from the adder, Vedral {et al.} provide an explicit construction of quantum circuit to realize modular exponentiation \cite{modular}. A plain adder is constructed in Fig. \ref{adder},  which consists of the CNOT gates and the Toffoli gates.

\begin{figure}[!h]
        \centering \includegraphics[width=1\columnwidth]{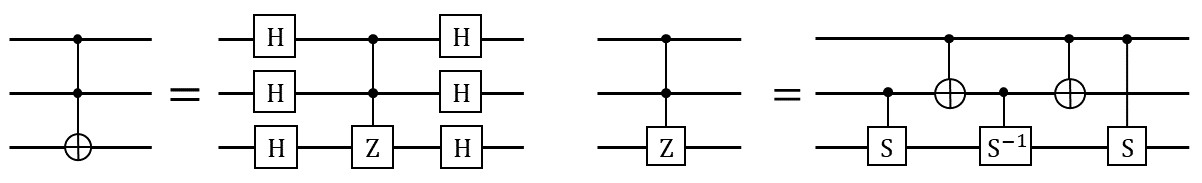}
        \caption{The constructions for Toffoli gate and $CCZ$.
        }\label{CCZ}
\end{figure}

The Toffoli gate is controlled-controlled-NOT and can be decomposed as
$$\rm{Toffoli=CCX={H}_{3}\cdot CCZ\cdot{H}_{3}}.$$
Fig. \ref{CCZ} shows one of constructions for the Toffoli gate, where $\rm CCZ=diag(1,1,1,1,1,1,1,-1)$ and ${H}_{3}=H\otimes H\otimes H$ with the Hadamard gate acting on the different 1-qubit.

\begin{figure}[!h]
        \centering \includegraphics[width=1\columnwidth]{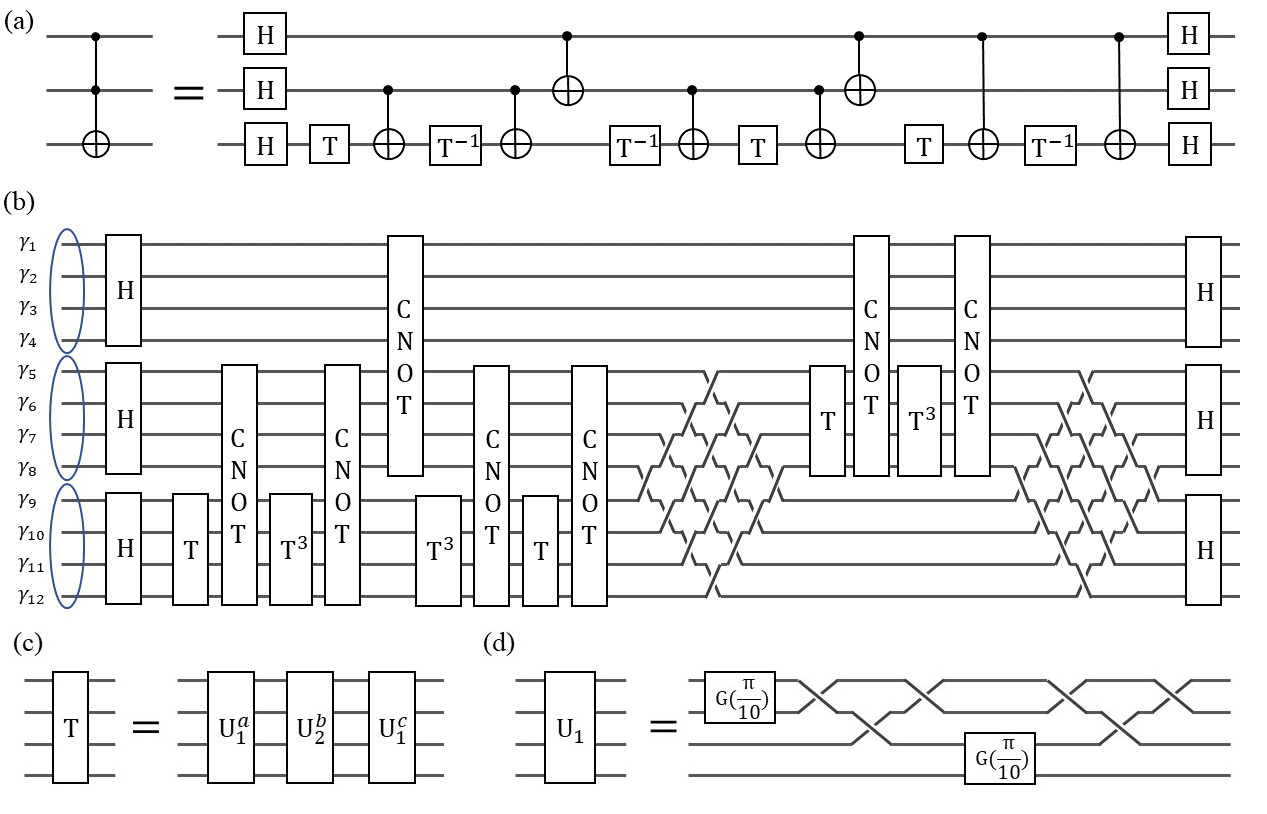}
        \caption{(a) The complete logical circuit for Toffoli gate. (b) The braiding diagram of (a).  (c) The braiding diagram of  the $T$-gate. (d) $\hat U_1$ operator. $\hat U_2$ can be constructed similarly.
        }\label{CCX}
\end{figure}

We draw the complete logical circuit for Toffoli gate in Fig. \ref{CCX} (a). And then Fig. \ref{CCX} (b) gives the braiding diagram with MEMs for the Toffoli gate. Notice that some quantum gates may act on two qubits which are not neighboring but  jumping is not allowed in the TQC. One has to move them together first. The operator $\hat U_1$ and $\hat U_2$ are composed by the Hadamard gate and different phase gates (see Section \ref{universal}). Fig. \ref{CCX}(d) shows the braiding diagram for $\hat U_1$ as an example.

\section{Physical realization of the quantum circuits}

\subsection{Hadamard Gate}

Figs. \ref{fig3} (a,b,d) and \ref{fig4} (a,b) are the schematics of the devices for 1-qubit and 2-qubits. Logically, these designed devices are equivalent to their corresponding world line graphs, i.e., Figs. \ref{fig3} (d) and \ref{fig4} (c). If we only measure these 1- or 2- qubits, we can use these devices. However, the devices designed in those schematics cannot be used as parts in the quantum circuit because the two-dimensional projects of the devices are topologically different from the world line graphs, which is easy to see by comparing Fig. \ref{fig3} (b) with (c), and Fig. \ref{fig4} (b) with (c).  Therefore, we change the designation of the devices in three dimensions {such} that they are adaptive with the world line graphs and then can be used to assemble the quantum circuit.

\begin{figure}[!h]
        \centering \includegraphics[width=1\columnwidth]{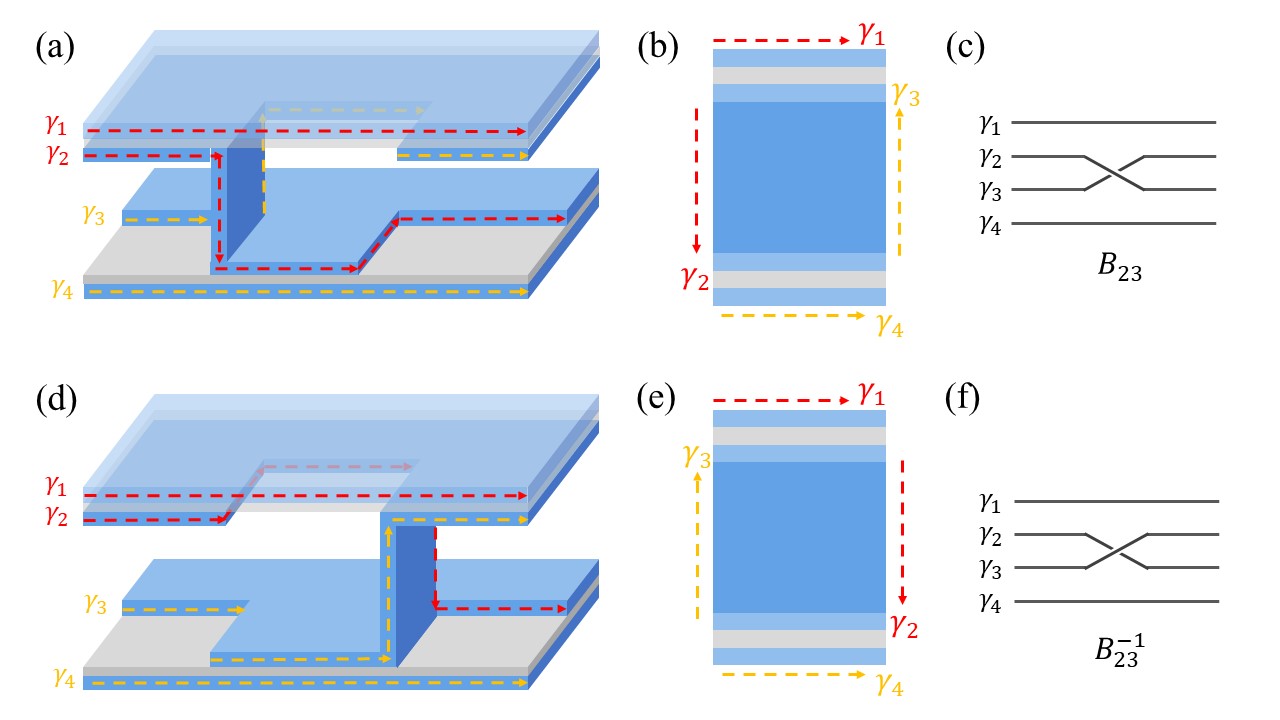}
        \caption{(Color online) (a)-(c) Device schematic of $B_{23}$ (as well as $B_{2n,2n+1}$). (d)-(f) Device schematic of $B_{23}^{-1}$. The blue layer is the $\chi$TSC thin films and gray layer is dielectric substrate. To be more intuitive, we make the upper layers translucent. The MEMs runs along the arrows in the edges of thin films.
        }\label{fB23}
\end{figure}

We first deform Fig. \ref{fig3} (a) for $B_{23}$ to Fig. \ref{fB23} (a) where the blue layer is the $\chi$TSC and the gray layer is dielectric substrate as before.  The second and the third $\chi$TSC layers are connected by a vertical $\chi$TSC slide and are corresponding to the $N=1$ base layer in Fig. \ref{fig3} (a). The top and down layers are corresponding to the two top layers on the right and left hand sides in Fig. \ref{fig3} (a).   The MEMs will run along the edges of $\chi$TSC and we see the braiding between $\gamma_{2,3}$.  Fig. \ref{fB23} (b) are right view of (a).  The front view of Fig. \ref{fB23} (a) maps out the world line (Fig. \ref{fB23}). The matrix $B_{23}$ in (\ref{B23}) is not  equal to its inverse. The inverse $B_{23}^{-1}$ is shown in Fig. \ref{fB23} (d). Its right and front views (Figs. \ref{fB23} (e) and (f)) indeed show that is an inverse operation.

It is also possible to exchange $\gamma_2$ and $\gamma_4$. However, for a 1-qubit gate, the probability of such an exchange is negligible because the low-energy pairing between two incoming fermions is forbidden by the superconducting gap.

\begin{figure}[!h]
        \centering \includegraphics[width=1\columnwidth]{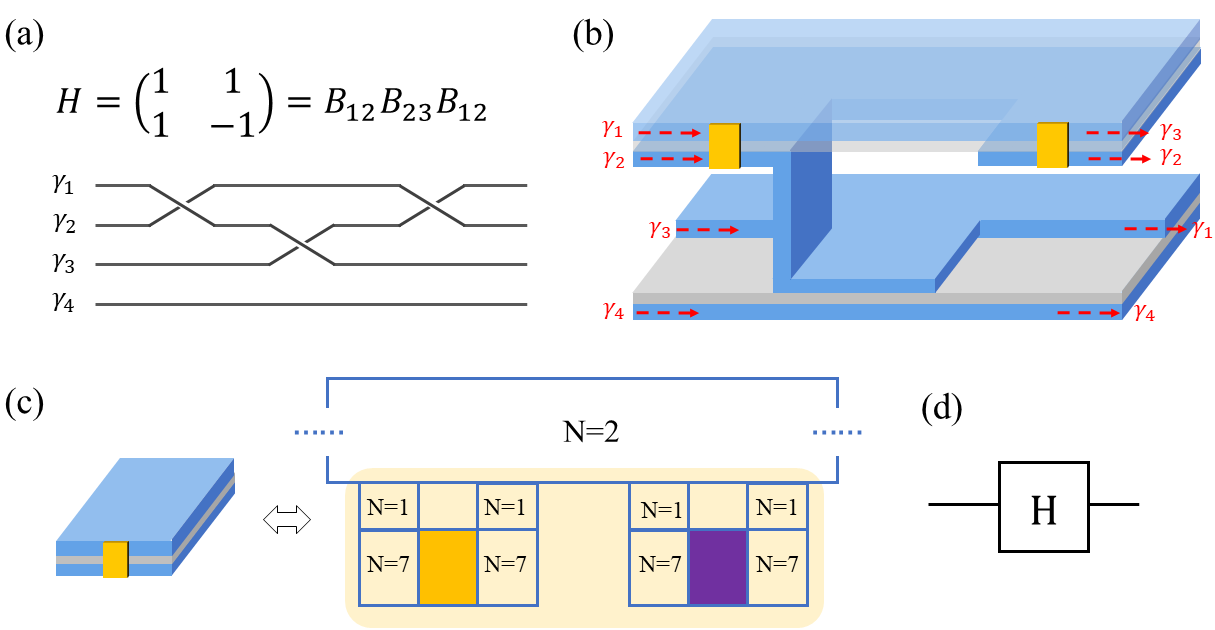}
        \caption{ (Color online) The construction of Hadamard gate. (a) The world line. (b) The device schematic. The yellow stickers are the phase gates. (c) The phase gate revisits. The yellow sticker on the left-hand side corresponds to the yellow shade area on the right-hand side, which is equivalent to Fig. \ref{fig2} (a) (d). The Hadamard gate in the quantum circuit model.
        }\label{H2}
\end{figure}

With the basic elements, the Hadamard gate can be constructed as Fig. \ref{H2}. The yellow stickers are the phase gates. 

\subsection{CNOT Gate}

\begin{figure}[!h]
        \centering \includegraphics[width=1\columnwidth]{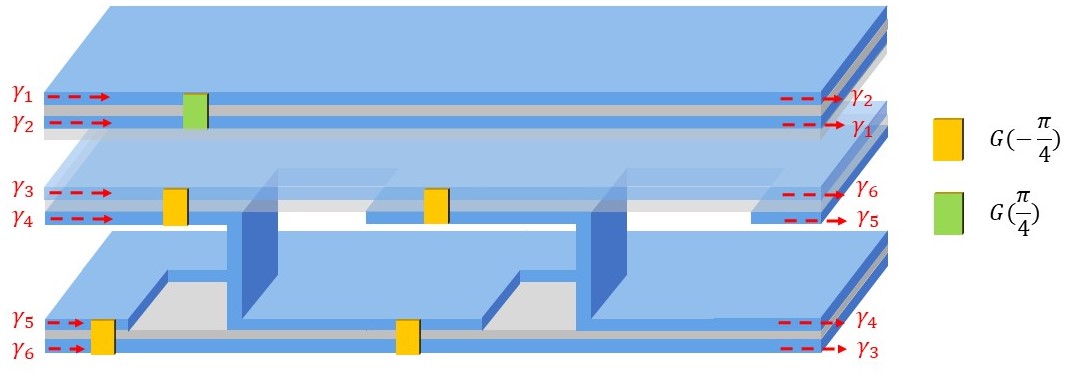}
        \caption{(Color online)  Braiding device of the CNOT in the dense encoding, corresponding to the blue box in Fig. \ref{S-CNOT1}. The green stickers are the $-\pi/4$ phase gates and the yellow {ones} are the inverse of the $\pi/4$ gates. It will be encapsulated together with two parity measurements corresponding to a logical CNOT gate in the sparse encoding.
        }\label{S-CNOT2}
\end{figure}

 Fig. \ref{S-CNOT2} shows the CNOT gate in the dense encoding and will be encapsulated together with two parity measurements corresponding to a logical CNOT gate as Fig. \ref{S-CNOT1}. In practical computation processes, if the whole circuit owns $n$ CNOT gates, we only keep the result that $\hat{P}_1=\hat{P}_2=...=\hat{P}_{2n}=1$ (or -1, depending on the selected FP). The probability of getting the useful results is $\frac{1}{4^{2n}}$. However, the speed that a quantum state runs is Fermi velocity $v_F$ which is $\sim10^{-3}c$. 
If the device size is in the order of microns,  we can input $10^{11}$ quantum states per second. 

As we mentioned in Sec. VII A,  we can alternatively choose to switch the CNOT$^{(+)}$ to the CNOT$^{(-)}$ after we get $i\gamma_4\gamma_5=-1$ in $M_1$. The  probability may increase to  $\frac{1}{2^{2n}}$. It is more efficient. 
The switching between $G(\pm\frac{\pi}4)$ and $G(\mp\frac{\pi}4)$ is operable because $G(\frac{\pi}4)=G(-\frac{\pi}4)G(-\frac{\pi}4)G(-\frac{\pi}4)$. The switching can be done by connecting either one $G(-\frac{\pi}4)$ or three $G(-\frac{\pi}4)$ after receiving the message  {in terms of} $M_1$. 

To further raise the inputing efficiency is possible instead of abandoning the wrong FP state measured in $M_2$. Some different process for measurement-based CNOT are proposed \cite{measure CNOT}. Instead of abandoning useless messages, they correct the quantum state according to the side measurements. The cost is 6 pairs of MEMs are needed and 
2 pairs of them are used to be ancillary and three side measurements are made. We can also realize this kind of measurement-based CNOT by the topological superconductor thin films. The advantage is the inputing efficient may greatly lift. How to correct the abandoning message in our algorithm is not studied here.


\section{ Conclusions }

 We proposed a new route to design a universal TQC. We showed that the universal TQC based on strongly corrected MEMs coincides with the conventional quantum circuit models. We have noticed that some quantum gates we designed are dependent on the FP of the input qubits. To construct the multi-qubits and the corresponding unitary transformation by braiding, we have used the sparse-dense mixed encoding process which is based on the FP side measurement to the ancillary qubits. We discussed the possibility to raise the inputing efficiency. We constructed a quantum circuit model for Shor's algorithm with our devices. In general, our process can be applied to all existed and new quantum algorithms in study \cite{rev}.
Without the FP measurement, we should follow the unitary category approach developed for the TQC \cite{TQCR}.   In this work, we do not do a full dense encoding construction for a universal TQC with our devices, which is a next task. We here take the strong coupling limit in which $\tau$ are reflected while the $\varepsilon$ are transmitted by interacting barrier. In reality, the reflection of anyons is not complete, but of a statistical correction as a recent experiment showed \cite{anyoncoll}. We will leave this effect from the anyon collisions to the TQC in a further study. We showed that some quantum operating results can transfer to the electrical signs of the output states, e.g., the CONT gate's in the dense encoding as well as the 1-qubit's. However, for a quantum circuit, because we take the sparse encoding and there are the other complex conditions, we do not discuss its electric sign of the output. We will  also study these issues in future. \\

 \centerline{\bf  Acknowledgements }
 
 \vspace{3mm}

 The authors thank Y. S. Wu for valuable discussions.
 This work is supported by NNSF of China with No. 12174067 (YMZ, BC,YY,XL), No. 11474061 (YMZ,YGC,YY,XL) and No. 11804223 (BC,XL) and the U.S. Department of Energy, Basic Energy Sciences Grant No. DE-FG02-99ER45747 (ZW).

\appendix

\section{Non-Abelian statistics of Majorana fermion approaches}\label{A}

We explain when the Majorana fermions are Abelian and when they are non-Abelian.

\subsection{One Species of Majorana Fermions}

Consider $N$-Majorana fermions which are of the same species, say, $\gamma(x_i)$, $i=1,...,N$, as in the $\nu=5/2$ FQHE and the Ising model \cite{MR}. Let us begin with the following Majorana fermion relations
\begin{eqnarray}
&&\gamma(x_i)\gamma(x_j)=- \gamma(x_j)\gamma(x_i),~ if~x_i\ne x_j\nonumber\\
&&\gamma(x_i)\gamma(x_j)=1, ~if~x_i=x_j\nonumber.
\end{eqnarray}
The quantum states at a given spatial point $x$ are $|\psi_+\rangle\oplus |\psi_-\rangle$ where $|\psi_\pm\rangle$ have fermion parity $\pm$. The fermion parity means whether the number of Majorana fermions is even or odd. The fusion rules are $ \psi_+\psi_+=\psi_+, \psi_+\psi_-=\psi_-, \psi_-\psi_-=\psi_+$, i.e., one species of many Majorana fermions are Abelian.

To enable non-Abelian statistics, one needs to introduce other degrees of freedom, e.g., vortices. If there is a vortex excitation besides the Majorana fermions at a given $x$, i.e., a Majorana bound state, we have the basis
\begin{eqnarray}
(|\psi_{+,0}\rangle,|\psi_{-,0}\rangle,|\psi_{+,1}\rangle,|\psi_{-,1}\rangle)\cong( I, \psi, \sigma, \sigma\times \psi=\mu).\nonumber\\ \label{1}
\end{eqnarray}
 where  $|\psi_{p,a}\rangle$ are the states with the fermion parity $p=\pm$ and the vortex number $0,1$. The right hand side of $\cong$ is the primary fields in the Ising model. According to the Ising model, the non-Abelian anyon is the Majorana bound state, 1 qubit, $\tilde\sigma=(\sigma+\mu,\sigma-\mu)$ which has the non-Abelian fusion rule $\tilde\sigma\times\tilde\sigma=I+\psi$.
 
 With anyon $\tilde\sigma$, one can construct the quantum gates $\{ H, \sqrt{Z}, CNOT\}$ and then TQC can be partially encoded.  

\subsection{Two Species of Majorana Fermions}

Now, let us discuss two species of Majorana fermions, which is the case in the main text,
where a spinless (or spin polarized) conventional fermion $\Psi$  can be decomposed into two species of Majorana fermions $\Psi(x_i)=\frac1{\sqrt 2}(\gamma^{(1)}(x_i)+i\gamma^{(2)}(x_i))$. $\Psi$s' anti-communication relations
    $$ \{\Psi(x_i),\Psi(x_j)\}=0, \{\Psi(x_i),\Psi^\dag(x_j)\}=\delta_{ij}$$
 require
\begin{eqnarray}
&&\gamma^{(1)}(x_i)\gamma^{(2)}(x_j)=-\gamma^{(1)}(x_j)\gamma^{(2)}(x_i), ~for ~any~x_i~and~x_j\nonumber\\
&&\gamma^{(a)}(x_i)\gamma^{(a)}(x_j)=-\gamma^{(a)}(x_j)\gamma^{(a)}(x_i),~ for~x_i\ne x_j\nonumber\\
&&\gamma^{(a)}(x_i)\gamma^{(a)}(x_j)=1, ~for~x_i=x_j\nonumber
\end{eqnarray}
Notice that two species of Majorana fermions are necessary to form a conventional (charged) fermion. If $\gamma^{(1)}\equiv\gamma^{(2)}$, i.e., they were the same species Majorana fermion, one could not obtain the above commutators for the conventional fermion.

For the conventional fermion in the normal states, the the particle number is conserved. The single particle state is $|\phi_n\rangle$ for  the particle number $n=0,1$.   We have the the well-known Fermi statistics which is Abelian.

In a superconductor, the single particle number is no longer a conserved quantity as a pair of electrons can turn into a Cooper pair and vice versa. However, the fermion parity, i.e., the odd/even number of the single particles, is conserved since the single particles are created or annihilated in pairs. Furthermore, in the edges of multi-layer thin films of a topological superconductor, two Majorana fermions from the injected conventional fermion can be delocalized to the edges of the different films (similar to two delocalized Majorana zero modes at the two ends of a Kitaev chain).  Thus, the quantum states at a position $x$ along the edge form the basis
\begin{eqnarray}
(|\varphi_{+,+}\rangle,|\varphi_{-,-}\rangle)\oplus(|\varphi_{+,-}\rangle,|\varphi_{-,+}\rangle)\equiv \Phi_e\oplus \Phi_o,\label{2}
\end{eqnarray}
where the subscripts $a,b=\pm$ in $\psi_{a,b}$ refer to the fermion parity of the first and second species of Majorana fermion.   $\Phi_{e,o}$ refers to the sector with the total fermion parity even/odd.  Each sector with a given total fermion parity is  two-fold degenerate. The fusion rule is $\Psi_o\times \Psi_o=\varphi_{+,+}+\varphi_{-,-}$.  Then being isomorphic to the Ising primary fields,  $\Psi_e\cong(I,\psi)$ while $\Psi_o\cong(\sigma, \mu)\cong \tilde\sigma$, which is the 1-qubit. Thus, we recover the fusion rule as that in the Ising model. Hence, the many-body Majorana  fermions of two species at the edge of multilayer thin films of the topological superconductor obey non-Abelian statistics.

This is also the case for the Kitaev chain \cite{K3}, the gapless Majorana edge modes of Kitaev honeycomb spin model \cite{K1}, and the model of Lian {et al.} \cite{BL}.

This two species approach has in fact been studied in earlier pioneering works, e.g.,  Nayak and Wilczek (Section 9 in \cite{NW})  and  Ivanov \cite{Inv}.

As we showed in this work, since we cannot braid the two Majorana objects from a charged fermion, we do not have $\sqrt{Z}$ gate. Thus, this type of non-Abelian Majorana objects cannot used to design a TQC because there is no a topological CNOT in this scheme. 

\section{ Basic facts of G$_2$}\label{B}

 Although Hu and Kane listed most of the useful contents of $G_2$ in their work \cite{hukane},
we would like to concisely repeat part of them for reader's convenience. The simplest exceptional Lie group
 G$_2$ as a subgroup of SO(7) keeps $\sum^7_{a,b,c=1} \hat{f}_{abc}\gamma_a\gamma_b\gamma_c$ invariant. We choose the nonzero total antisymmetric $\hat{f}_{abc}$ to be \cite{G2}
 \begin{eqnarray}
\hat{f}_{124}=\hat{f}_{235}=\hat{f}_{346}=\hat{f}_{457}=\hat{f}_{561}=\hat{f}_{672}=\hat{f}_{713}=1,\nonumber\\
\end{eqnarray}
and the permutations. The 21 generators of SO(7) can be represented by $7\times7$ skew matrices $L^{m,n}_{ab}=i(\delta_{ma}\delta_{nb}-\delta_{na}\delta_{mb})$ where $m<n=1,...,7$. The dimensions of $G_2$ is 14 and the generators  $\Xi^\alpha$ of the fundamental representation of G$_2$ is given by \cite{hukane,GF}
 \begin{eqnarray}
 \Xi^\alpha=\begin{cases}
 \frac{L^{\alpha,\alpha+2}-L^{\alpha+1,\alpha+5}}{\sqrt2},&\alpha=1,...,7\\
 \frac{L^{\alpha,\alpha+2}+L^{\alpha+1,\alpha+5}-2L^{\alpha+3,\alpha+4}}{\sqrt6},&\alpha=8,...,14.
 \end{cases}
 \end{eqnarray}
 The quadratic Casimir operator is given by
 \begin{eqnarray}
 \sum_{\alpha=1}^{14}\Xi^\alpha_{ab}\Xi^\alpha_{cd}=\frac{2}3(\delta_{ad}\delta_{bc}-\delta_{ac}\delta_{bd})-\frac{1}{18}\sum_{efg}\epsilon_{abcdefg}\hat{f}_{efg},\nonumber\\
 \label{qc}
  \end{eqnarray}
where $\epsilon_{abcdefg}$ is the 7-dimensional total antisymmetric tensor.  	
	
\section{Details of the interaction terms}\label{C}
The interactions in the main text read
 \begin{eqnarray}
 H_i=-\frac{\lambda}3\sum_{a\ne b}\gamma^R_a\gamma^R_b\gamma^L_b\gamma^L_a-\frac{\lambda}3{\sum}'\gamma^R_a\gamma^R_b\gamma^L_c\gamma^L_d, \label{int1a}
  \end{eqnarray}
where $\sum'$ means the summation runs the indices with $\epsilon_{abcdefg}\hat{f}_{efg}=-1$.  In Fig. \ref{fig5}, we give an example of the interaction domains, i.e., $\gamma_3^L\gamma_5^L\gamma_6^R\gamma_7^R$. There are 42 terms in the first sum of Eq. \ref{int1a} and half of them are non-equivalent. There are also 42 terms in the second sum. We list all of them as follows:
\begin{widetext}
\begin{eqnarray}
&&\sum_{a\ne b}\gamma^R_a\gamma^R_b\gamma^L_b\gamma^L_a=2\sum_{a< b}\gamma^R_a\gamma^R_b\gamma^L_b\gamma^L_a\nonumber\\
&&=2(\gamma^R_1\gamma^R_2\gamma^L_2\gamma^L_1+\gamma^R_1\gamma^R_3\gamma^L_3\gamma^L_1+\gamma^R_1\gamma^R_4\gamma^L_4\gamma^L_1+\gamma^R_1\gamma^R_5\gamma^L_5\gamma^L_1+\gamma^R_1\gamma^R_6\gamma^L_6\gamma^L_1+\gamma^R_1\gamma^R_7\gamma^L_7\gamma^L_1+\gamma^R_2\gamma^R_3\gamma^L_3\gamma^L_2\nonumber\\
&&+\gamma^R_2\gamma^R_4\gamma^L_4\gamma^L_2+\gamma^R_2\gamma^R_5\gamma^L_5\gamma^L_2+\gamma^R_2\gamma^R_6\gamma^L_6\gamma^L_2+\gamma^R_2\gamma^R_7\gamma^L_7\gamma^L_2+\gamma^R_3\gamma^R_4\gamma^L_4\gamma^L_3+\gamma^R_3\gamma^R_5\gamma^L_5\gamma^L_3+\gamma^R_3\gamma^R_6\gamma^L_6\gamma^L_3\nonumber\\
&&+\gamma^R_3\gamma^R_7\gamma^L_7\gamma^L_3+\gamma^R_4\gamma^R_5\gamma^L_5\gamma^L_4+\gamma^R_4\gamma^R_6\gamma^L_6\gamma^L_4+\gamma^R_4\gamma^R_7\gamma^L_7\gamma^L_4+\gamma^R_5\gamma^R_6\gamma^L_6\gamma^L_5+\gamma^R_5\gamma^R_7\gamma^L_7\gamma^L_5+\gamma^R_6\gamma^R_7\gamma^L_7\gamma^L_6),\\
&&{\sum}'\gamma^R_a\gamma^R_b\gamma^L_c\gamma^L_d\nonumber\\
&&=\gamma^R_3\gamma^R_5\gamma^L_6\gamma^L_7+\gamma^R_3\gamma^R_6\gamma^L_5\gamma^L_7+\gamma^R_3\gamma^R_7\gamma^L_5\gamma^L_6+\gamma^R_6\gamma^R_7\gamma^L_3\gamma^L_5+\gamma^R_7\gamma^R_5\gamma^L_3\gamma^L_6+\gamma^R_5\gamma^R_6\gamma^L_3\gamma^L_7\nonumber\\
&&+\gamma^R_4\gamma^R_6\gamma^L_7\gamma^L_1+\gamma^R_4\gamma^R_7\gamma^L_1\gamma^L_6+\gamma^R_4\gamma^R_1\gamma^L_6\gamma^L_7+\gamma^R_7\gamma^R_1\gamma^L_4\gamma^L_6+\gamma^R_1\gamma^R_6\gamma^L_4\gamma^L_7+\gamma^R_6\gamma^R_7\gamma^L_4\gamma^L_1\nonumber\\
&&+\gamma^R_5\gamma^R_7\gamma^L_1\gamma^L_2+\gamma^R_5\gamma^R_1\gamma^L_2\gamma^L_7+\gamma^R_5\gamma^R_2\gamma^L_7\gamma^L_1+\gamma^R_1\gamma^R_2\gamma^L_5\gamma^L_7+\gamma^R_2\gamma^R_7\gamma^L_5\gamma^L_1+\gamma^R_7\gamma^R_1\gamma^L_5\gamma^L_2\nonumber\\
&&+\gamma^R_6\gamma^R_1\gamma^L_2\gamma^L_3+\gamma^R_6\gamma^R_2\gamma^L_3\gamma^L_1+\gamma^R_6\gamma^R_3\gamma^L_1\gamma^L_2+\gamma^R_2\gamma^R_3\gamma^L_6\gamma^L_1+\gamma^R_3\gamma^R_1\gamma^L_6\gamma^L_2+\gamma^R_1\gamma^R_2\gamma^L_6\gamma^L_3\nonumber\\
&&+\gamma^R_7\gamma^R_2\gamma^L_3\gamma^L_4+\gamma^R_7\gamma^R_3\gamma^L_4\gamma^L_2+\gamma^R_7\gamma^R_4\gamma^L_2\gamma^L_3+\gamma^R_3\gamma^R_4\gamma^L_7\gamma^L_2+\gamma^R_4\gamma^R_2\gamma^L_7\gamma^L_3+\gamma^R_2\gamma^R_3\gamma^L_7\gamma^L_4\nonumber\\
&&+\gamma^R_1\gamma^R_3\gamma^L_4\gamma^L_5+\gamma^R_1\gamma^R_4\gamma^L_5\gamma^L_3+\gamma^R_1\gamma^R_5\gamma^L_3\gamma^L_4+\gamma^R_4\gamma^R_5\gamma^L_1\gamma^L_3+\gamma^R_5\gamma^R_3\gamma^L_1\gamma^L_4+\gamma^R_3\gamma^R_4\gamma^L_1\gamma^L_5\nonumber\\
&&+\gamma^R_2\gamma^R_4\gamma^L_5\gamma^L_6+\gamma^R_2\gamma^R_5\gamma^L_6\gamma^L_4+\gamma^R_2\gamma^R_6\gamma^L_4\gamma^L_5+\gamma^R_5\gamma^R_6\gamma^L_2\gamma^L_4+\gamma^R_6\gamma^R_4\gamma^L_2\gamma^L_5+\gamma^R_4\gamma^R_5\gamma^L_2\gamma^L_6.
\end{eqnarray}
\begin{figure}
	\includegraphics[width=0.6\textwidth]{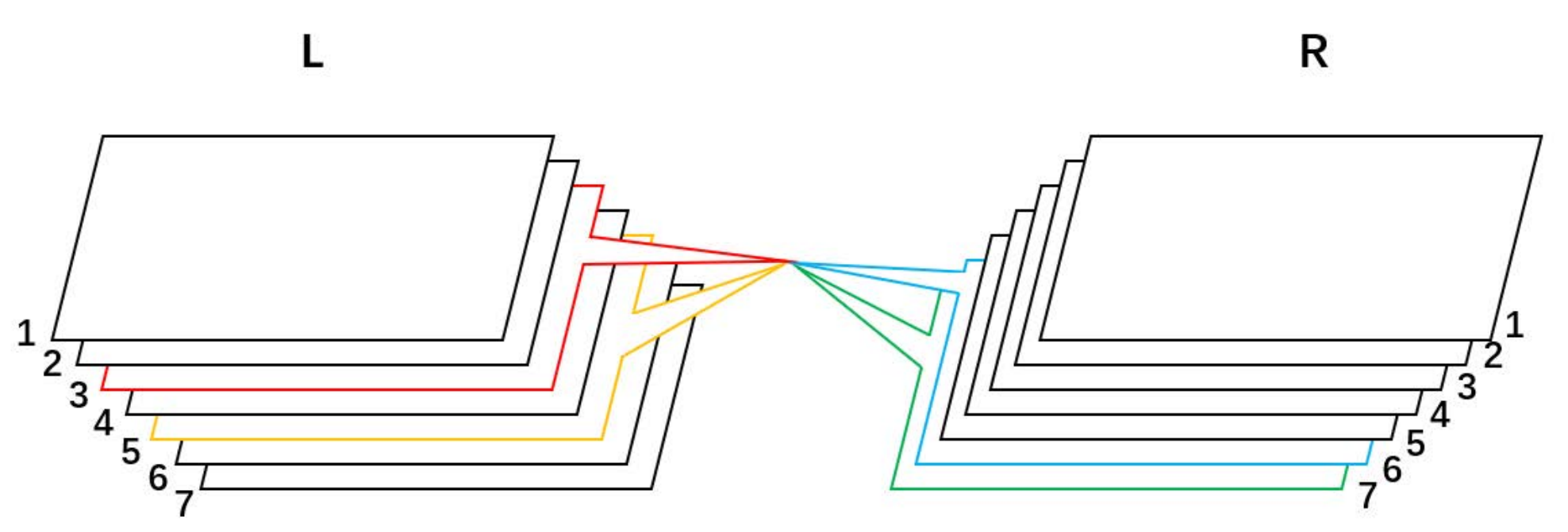}
	\caption{(Color online)
		The illustrations of the interactions $\gamma_3^L\gamma_5^L\gamma_6^R\gamma_7^R$. Each layer provides a $\chi$MEM channel with Chern number $N=1$. The detailed illustrations at the joint of Layer 3L, 5L, 6R, and 7R are shown in Fig. 1(b) and (d) in the main text.
		\label{fig5}	}
\end{figure}
\end{widetext}

\section{details of device realizations of $G(-\frac{\pi}{4})$, $G(\frac{\pi}{10})$ and $G(\frac{2\pi}{5})$}\label{D}

\begin{widetext}	
	As described in the main text, by a proper arrangement of the interaction areas, the device in Fig. 3(a) of the main text can realize $G(-\frac{\pi}{4})$, $G(\frac{\pi}{10})$ and $G(\frac{2\pi}{5})$. The details are shown in Fig. \ref{fig6}.
	
	  \begin{figure}
	  	\centerline{\includegraphics[width=0.6\textwidth]{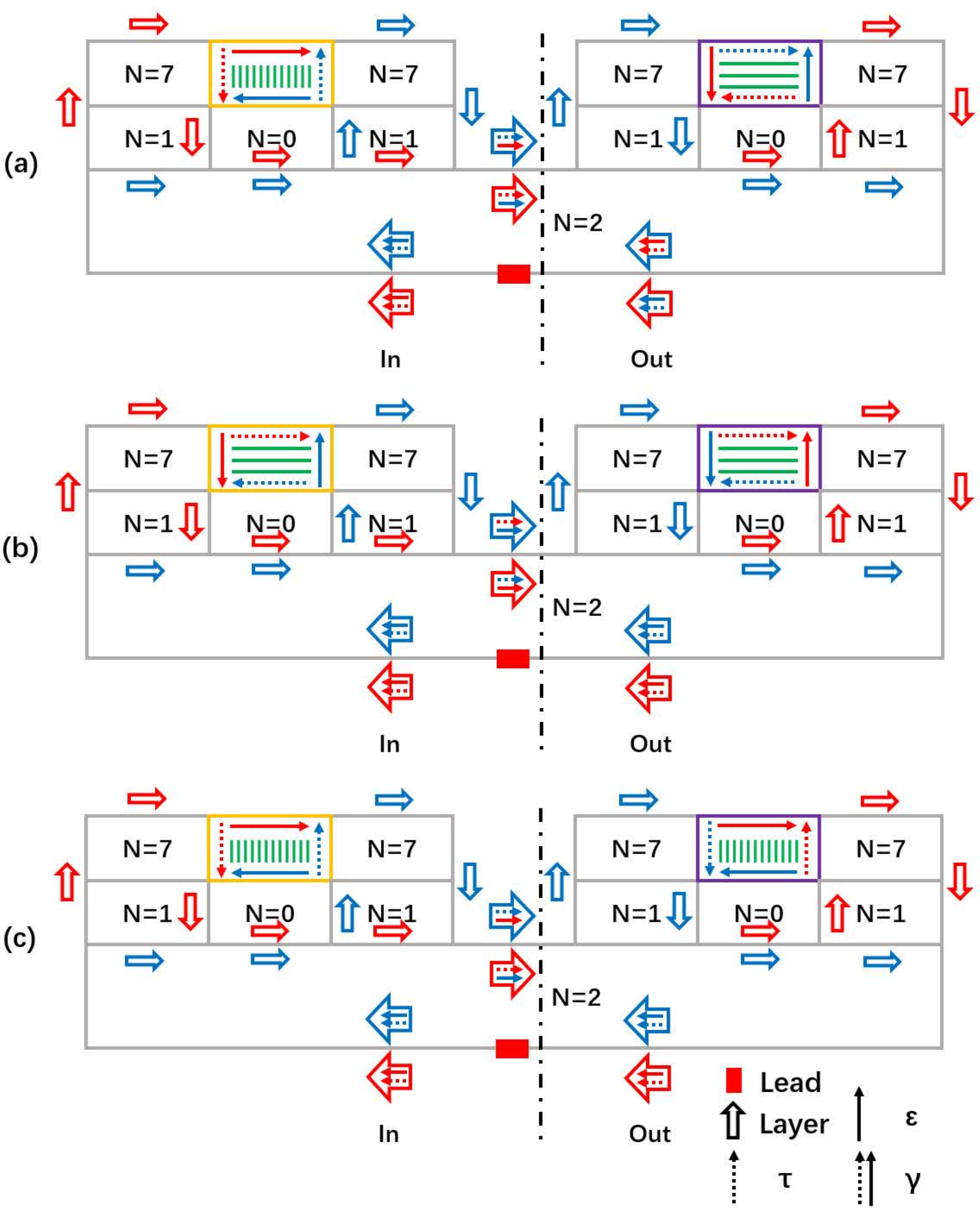}}
	  	\caption{(Color online) (a), (b), and (c) are the top view of the device for $G(-\frac{\pi}{4})$, $G(\frac{\pi}{10})$ and $G(\frac{2\pi}{5})$ in Fig. 3 of the main text.  \label{fig6}}
	  \end{figure}
\end{widetext}

\section{The electric signals with phase gates} \label{F}	

Consider the setup in Fig. 4 in the main text. The phase gate is represented by $G_{ij}(\theta)={\rm diag}(1, e^{-i2\theta})={\rm diag}(1, e^{-i\eta}) $ on the basis $|0^{\gamma_i\gamma_j}\rangle$ and $|1^{\gamma_i\gamma_j}\rangle$, where $0$ and $1$ labels the fermion number. Then,  in the parity even basis {$(|0_A0_B0_C\rangle, |0_A1_B1_C\rangle, |1_A0_B1_C\rangle, |1_A1_B0_C\rangle)^T$}, the transformation matrix corresponding to Fig. 4 in the main text reads
\begin{widetext}
\begin{equation}
\frac{1}{2}
\left(
\begin{array}{cccc}
-(1-e^{-i \text{$\eta_{5}$}}) & ie^{-i \text{$\eta_{12}$}} (1+e^{-i \text{$\eta_{5}$}}) & 0 & 0 \\
i(1+e^{-i \text{$\eta_{5}$}}) & e^{-i \text{$\eta_{12}$}} (1-e^{-i \text{$\eta_{5}$}}) & 0 & 0 \\
0 & 0 & e^{-i \text{$\eta_{13}$}} \left(-e^{-i \text{$\eta_{4}$}}+e^{-i \text{$\eta_{45}$}}\right) & ie^{-i \text{$\eta_{23}$}} \left(e^{-i \text{$\eta_{4}$}}+e^{-i \text{$\eta_{45}$}}\right) \\
0 & 0 & ie^{-i \text{$\eta_{13}$}} \left(e^{-i \text{$\eta_{4}$}}+e^{-i \text{$\eta_{45}$}}\right) & e^{-i \text{$\eta_{23}$}} \left(e^{-i \text{$\eta_{4}$}}-e^{-i \text{$\eta_{45}$}}\right) \\
\end{array}
\right)
\end{equation}
\end{widetext}
where $\eta_{ij}=\eta_i+\eta_j$ and $\eta_{ijk}=\eta_i+\eta_j+\eta_k$. Notice that the conductance between Lead 2 and {Lead} 3 is $\sigma_{23}=(1-|\langle\psi_f|\psi_i\rangle|^2)e^2/h$, then if we choose the initial state $|\psi_i\rangle=|0_A0_B1_C\rangle$, then the final state is
\begin{eqnarray}
|\psi_f\rangle&=&-\frac{1}{2}(1-e^{-i \text{$\eta_{5}$}})|0_A0_B0_C\rangle\nonumber\\
&+&\frac{1}{2}[ie^{-i \text{$\eta_{12}$}} (1+e^{-i \text{$\eta_{5}$}})]|0_A1_B1_C\rangle,
\end{eqnarray}
and the corresponding conductance is {$\sigma_{23}=\cos^2(\theta_5)\frac{e^2}{h}$}. Similar results for electric signals can be obtained for the other three initial states. Furthermore, one solution for the CNOT gate is { $2\theta_1=\frac{\pi}{2}$ and $2\theta_2=2\theta_3=2\theta_4=2\theta_5=-\frac{\pi}{2}$,} and the they can be realized through topological $\frac{\pi}{4}$-phase gates.

\section{Details of the CNOT gate in sparse encoding}\label{CNOTdetial}	

In the sparse encoding, 2-qubits are associated with 4 pairs of MEMs $(\gamma_1,\gamma_2),...,(\gamma_7,\gamma_8)$ labeled as $A,B,C,D$. If choosing the even FP input state given by ${\gamma}_{1}{\gamma}_{2}{\gamma}_{3}{\gamma}_{4}={\gamma}_{5}{\gamma}_{6}{\gamma}_{7}{\gamma}_{8}=+1 $, the basis of system is
$$\{|0_A0_B0_C0_D\rangle,|0_A0_B1_C1_D\rangle,|1_A1_B0_C0_D\rangle,|1_A1_B1_C1_D\rangle\}.$$

We take $\gamma_{4,5}$ as an ancillary qubit and  measure its FP, i.e., $i{\gamma}_{4}{\gamma}_{5}$. Before measuring, $\gamma_{6}$ exchanges with $\gamma_{4,5}$ and forms a new pair with $\gamma_{3}$. This equals to act $B_{45}B_{56}$ on the $ {\gamma}_{3}{\gamma}_{4}{\gamma}_{5}{\gamma}_{6} $ and gives superposition states
\begin{equation}
\left(
\begin{array}{ccc}
|0_A0_B0_C0_D\rangle+|0_A1_B1_C0_D\rangle \\
i|0_A0_B1_C1_D\rangle-i|0_A1_B0_C1_D\rangle \\
|1_A0_B1_C0_D\rangle+|1_A1_B0_C0_D\rangle \\
-i|1_A0_B0_C1_D\rangle+i|1_A1_B1_C1_D\rangle \\
\end{array}
\right)
\end{equation}

We then measure $i{\gamma}_{4}{\gamma}_{5}$ and go ahead when it is even. This means that the fermion number of the pair $C$ is 0, and the remaining state is $$\{|0_A0_B0_D\rangle,i|0_A1_B1_D\rangle,|1_A1_B0_D\rangle,i|1_A0_B1_D\rangle\},$$
which is 2-qubits for the dense encoding in odd FP.

Acting the CNOT in the dense encoding on this state, the output state is given by
$$\{|0_A0_B0_D\rangle,i|0_A1_B1_D\rangle,|1_A0_B1_D\rangle,i|1_A1_B0_D\rangle\}$$
Putting $\gamma_{4,5}$ with $i{\gamma}_{4}{\gamma}_{5}=1$ back, we have
$$\{|0_A0_B0_C0_D\rangle,i|0_A1_B0_C1_D\rangle,|1_A0_B0_C1_D\rangle,i|1_A1_B0_C0_D\rangle\}.$$
Braiding $\gamma_{6}$ with ${\gamma}_{4},{\gamma}_{5}$ again,  we get the superposition state
\begin{equation}
\left(
\begin{array}{ccc}
|0_A0_B0_C0_D\rangle-i|0_A1_B1_C0_D\rangle \\
|0_A0_B1_C1_D\rangle-i|0_A1_B0_C1_D\rangle \\
i|1_A0_B0_C1_D\rangle+|1_A1_B1_C1_D\rangle \\
i|1_A0_B1_C0_D\rangle+|1_A1_B0_C0_D\rangle \\
\end{array}
\right)
\end{equation}

Finally we measure ${\gamma}_{5}{\gamma}_{6}{\gamma}_{7}{\gamma}_{8}$. If it is $+1$, we have the FP even output state in the sparse encoding
$$\{|0_A0_B0_C0_D\rangle,|0_A0_B1_C1_D\rangle,|1_A1_B1_C1_D\rangle,|1_A1_B0_C0_D\rangle\}$$
This process implements a CNOT gate in the sparse encoding with the FP even. Similarly,  we can also design the CNOT gate with FP odd.

\end{document}